\documentclass[a4paper,fleqn,usenatbib]{mnras}

\usepackage{newtxtext,newtxmath}

\usepackage[T1]{fontenc}
\usepackage{ae,aecompl}

\usepackage{graphicx}	
\usepackage{amsmath}	
\usepackage{amssymb}	
\usepackage{textcomp}   

\usepackage[capitalise]{cleveref}

\title[C/O ratios of exoplanet atmospheres]{The carbon-to-oxygen ratio: implications for the spectra of hydrogen-dominated exoplanet atmospheres}

\author[Drummond et al.]{
Benjamin Drummond$^{1}$\thanks{E-mail: b.drummond@exeter.ac.uk},
Aarynn L. Carter$^{1}$,
Eric H\'ebrard$^{1}$,
Nathan J. Mayne$^{1}$,
\newauthor
David K. Sing$^{2}$,
Thomas M. Evans$^{3}$,
and Jayesh Goyal$^{1}$
\\
$^{1}$Astrophysics Group, University of Exeter, Exeter, EX4 4QL, UK\\
$^{2}$Department of Earth and Planetary Sciences, Johns Hopkins University, Baltimore, MD, USA\\
$^{3}$Kavli Institute for Astrophysics and Space Research, Massachusetts Institute of Technology, 77 Massachusetts Avenue, 37-241, Cambridge, MA 02139, USA
}

\date{Accepted 2019 March 25. Received 2019 February 15; in original form 2018 December 20}

\pubyear{2019}

\begin{document}
\label{firstpage}
\pagerange{\pageref{firstpage}--\pageref{lastpage}}
\maketitle

\begin{abstract}

We present results from one-dimensional atmospheric simulations investigating the effect of varying the carbon-to-oxygen (C/O) ratio on the thermal structure, chemical composition and transmission and emission spectra, for irradiated hydrogen-dominated atmospheres. We find that each of these properties of the atmosphere are strongly dependent on the individual abundances (relative to hydrogen) of carbon and oxygen. We confirm previous findings that different chemical equilibrium compositions result from different sets of element abundances but with the same C/O ratio. We investigate the effect of this difference in composition on the thermal structure and simulated spectra. We also simulate observations using the PandExo tool and show that these differences are observationally significant with current (i.e. Hubble Space Telescope) and future (i.e. James Webb Space Telescope) instruments. We conclude that it is important to consider the full set of individual element abundances, with respect to hydrogen, rather than the ratios of only two elements, such as the C/O ratio, particularly when comparing model predictions with observed transmission and emission spectra. 

\end{abstract}



\section{Introduction}
\label{section:intro}

The carbon-to-oxygen (C/O) ratio and its effect on the atmospheric properties and observable spectra of transiting exoplanets has been well studied \citep[e.g.][]{SeaRH05,Lod10,MadHS11,Moses2013}. It has been suggested that the C/O ratio could be a fundamental parameter for classifying exoplanet atmospheres due to its potential to link the current atmospheric chemical composition with the formation and evolution of the planet \citep[e.g.][]{Gai00,KucS05,Mad12}. It is known that the C/O ratio varies across the stellar population \citep[e.g.][]{DelIG10,PetM11,Nis13,TesCS14,BreF16} and that the C/O ratio of planets can depart from that of their host stars due to complex processes during the formation and evolution of a planet within a dynamically, thermally and chemically active proto-planetary disk \citep[e.g.][]{ObeMB11,PisOB15,MorvM16,ObeB16,BreFM17,EspFM17,EisWv18}.

An obvious transition occurs at a C/O ratio of unity: for C/O $>1$ carbon is more abundant than oxygen but for C/O $<1$ carbon is less abundant than oxygen. This important transition has been the focus of many theoretical studies of exoplanet atmospheres \citep[e.g.][]{ForML05,MadMJ11,Kopparapu2012,Moses2013,MosLV13,Venot2015,MolvD15,HenL16,TsaLG2017,GanM17,GoyMS18}. The C/O ratio has also been considered using inverse, retrieval models that retrieve the atmospheric properties which best fit a given set of observations, resulting in a wide-range of reported C/O values for transiting exoplanets \citep[e.g.][]{MadS09,KonBM13,LinWZ13,LinKW14,LeeHI13,SteBM14,WalTR15,KreLB15,LavMM17,WakSD18,MarFS18,GanM18}.

A common starting point for chemical models of hydrogen-dominated exoplanet atmospheres is the assumption of a set of element abundances corresponding to the solar composition. The effect of assuming a different set of element abundances on the structure and properties of these atmospheres has been well investigated, typically in one of two approaches. Firstly, a number of studies \citep[e.g.][]{ForML05,LewSF2010,AguVS2014} have investigated the effect of scaling the abundances of the heavy elements (heavier than hydrogen and helium) uniformly, such that the abundance ratios of the heavy elements with each other are fixed. Secondly, other studies have focussed on the effect of varying the abundance ratios of the heavy elements themselves, primarly the C/O ratio \citep[e.g.][]{MadMJ11,MosLV13,Venot2015} but also the N/O ratio \citep{Mig19}.

Starting with some initial set of element abundances the C/O ratio could be increased either by increasing the carbon abundance (C/H, relative to hydrogen) or decreasing the oxygen abundance (O/H, relative to hydrogen). Likewise, the C/O ratio could be decreased either by decreasing C/H or increasing O/H. Clearly, the C/O ratio alone is not enough to specify the full set of element abundances and hence the chemical equilibrium composition, for a given pressure and temperature. For example, \citet[][Fig. 11]{MosLV13} show that for a given C/O ratio the mole fractions of many key chemical species are strongly dependent on the values of C/H and O/H.

A number of previous studies investigating the effect of the C/O ratio on the atmospheric structure clearly state that they assume a fixed value for O/H and vary C/H \citep{HelL09,MadMJ11,Moses2013,TsaLG2017}, while others choose to fix the value of C/H and vary O/H \citep{MosLV13,MolvD15,GoyMS18}.  \citet{KitHR18} investigate the dependence of the carbon monoxide (CO) mole fraction on the C/O ratio both by varying C/H and O/H, with the other held fixed. \citet{HenL16} and \citet{HenT16} consider the equilibrium composition for two different sets of element abundances with the same C/O ratio but different C/H and O/H. We note that some studies do not explicitly state how they change their set of element abundances to achieve different C/O ratios.

In this paper we investigate the effect of different sets of element abundances on the structure and predicted spectra of hot Jupiter and warm Neptune atmospheres. We focus on the differences that can result from assuming sets of element abundances that have the same C/O ratio, but with different C/H and O/H. We find a significant difference in the chemical and thermal structure, as well as the predicted spectra between models with the same C/O ratio but different combinations of O/H and C/H. In \cref{section:desc} we describe our numerical model and methodology. In \cref{section:forward} we quantify the impact of different methods of varying the C/O ratio on the temperature structure, chemical composition and spectra, using a 1D atmosphere model. In \cref{section:discussion} we discuss some implications of our results and briefly review previous studies relating to the bulk chemical composition of exoplanets, focusing on the C/O ratio. Finally, we present our conclusions in \cref{section:conclusions}.

\section{Model Description and methods}
\label{section:desc}
\subsection{1D atmosphere model: ATMO}

We use the 1D atmosphere model ATMO \citep{Tremblin2015,Tremblin2016,DruTB16,GoyMS18} to calculate pressure--temperature ($P$-$T$) profiles, the chemical equilibrium composition and emission and transmission spectra for hydrogen-dominated, irradiated exoplanet atmospheres. A description of the model can be found in \citet{DruTB16} and \citet{GoyMS18}, though we summarise the key points here. 

The model solves for hydrostatic balance and radiative--convective equilibrium, with an internal heat flux and irradiation at the top of the atmosphere as boundary conditions. The radiative transfer equation is solved in 1D plane-parallel geometry and includes isotropic scattering. We include CH$_4$, H$_2$O, CO, CO$_2$, NH$_3$, HCN, C$_2$H$_2$, Na, Li, Rb and Cs, as well as H$_2$-H$_2$ and H$_2$-He collision--induced absorption, as opacity sources. We refer the reader to \citet{GoyMS18} for the most up-to-date sources for the line lists. The model uses the correlated--$k$ approximation with the method of random overlap to calculate the combined opacity of the mixture \citep{Lacis1991,AmuTM17}. To compute the radiative flux in the radiative-convective equilibrium iterations we use 32 bands while for the synthetic spectra we use 5000 bands, to achieve a higher spectral resolution. The chemical equilibrium composition is solved for using a Gibbs energy minimisation scheme based on the method of \citet{Gordon1994}.

In \cref{section:1d} and \cref{section:spectra} we investigate two types of hydrogen-dominated atmosphere: a ``hot'' case where CO is the most stable form of carbon and a ``warm'' case where methane (CH$_4$) is the most stable form of carbon. The model parameters for both of these cases are shown in \cref{table:param}. The parameters for the hot and warm cases are loosely based on HD~209458b and GJ~436b, respectively, though our aim is not to accurately model these specific planets. For the irradiation spectra, we use the Kurucz spectrum of HD~209458\footnote{\url{http://kurucz.harvard.edu/stars.html}} for the hot atmosphere and a Phoenix BT--Settl model spectrum \citep[][]{Allard2012}, with parameters $T_{\rm eff} = 3700$ K, $\log g = 5$, [M/H] = -0.3, for the warm atmosphere. 

\begin{table} 
\caption{Model parameters for the hot and warm atmospheres}
\label{table:param}
\renewcommand{\footnoterule}{}  
\begin{tabular}{l c c}
\hline\hline  
Parameter & Hot atmosphere & Warm atmosphere \\
\hline  
Surface gravity [ms$^{-1}$] & 10 & 10 \\
Planet radius [R$_{\rm J}$] & 1.0 & 0.35 \\
Intrinsic temperature [K] & 100 & 100 \\
Stellar radius [R$_{\odot}$] & 1.0 & 0.45 \\
Semi-major axis [AU] & 0.04 & 0.03 \\
Zenith angle [$\cos\theta$] & 0.5 & 0.5 \\
Redistribution parameter & 0.5 & 0.5 \\
\hline
\end{tabular}
\end{table}

\subsection{Element abundances}
\label{section:ele}

\begin{table*} 
\caption{Element abundance ratios for the solar photosphere \citep{Asplund2009} and for different sets of element abundances with C/O = 0.1 and C/O = 2 that vary C/H or O/H, with other element held fixed to the solar value, as used in \cref{section:1d,section:spectra}}
\label{table:ratios}
\renewcommand{\footnoterule}{}  
\begin{tabular}{l c c c c c}
\hline\hline  
& Solar & C/O = 0.1 & C/O = 0.1 & C/O = 2 & C/O = 2\\
& \citep{Asplund2009} & (vary O/H, fix C/H) & (vary C/H, fix O/H) & (vary O/H, fix C/H) & (vary C/H, fix O/H) \\
\hline  
C/H & $2.69\times10^{-4}$ & $2.69\times10^{-4}$ & $4.90\times10^{-5}$ & $2.69\times10^{-4}$ & $9.80\times10^{-4}$ \\
O/H & $4.90\times10^{-4}$ & $2.69\times10^{-3}$ & $4.90\times10^{-4}$ & $1.35\times10^{-4}$ & $4.90\times10^{-4}$   \\
N/H & $6.76\times10^{-5}$ & $6.76\times10^{-5}$ & $6.76\times10^{-5}$ & $6.76\times10^{-5}$ & $6.76\times10^{-5}$  \\
\hline
C/O & 0.55 & 0.10 & 0.10 & 2.00 & 2.00 \\
C/N & 3.98 & 3.98 & 0.72 & 3.98 & 14.50  \\
O/N & 7.25 & 39.79 & 7.25 & 2.00 & 7.25  \\
\hline
\end{tabular}
\end{table*}

An important input to the model are the set of element abundances which, along with the pressure and temperature, determine the chemical equilibrium composition. In this paper we choose the solar photospheric abundances of \citet{Asplund2009} for our ``baseline'' set of element abundances. The solar photospheric abundances are inferred by comparing the observed solar spectrum with predictions derived from solar atmosphere models \citep[see][for a review]{Pri16}. Reported abundances are frequently revised in light of new model developments and abundances can vary significantly between studies \citep[e.g.][]{GreS98,Lodders2009,Asplund2009,Caffau2011}. Choosing to use a different set of solar element abundances will likely result in quantitative differences in our results but will not likely effect the qualitative trends that we find.

In \cref{section:box,section:mmm} we investigate the effect of varying the C/O ratio on the chemical composition, for different metallicities. For calculations where we define O/H by the C/O ratio we uniformly scale the abundances of all elements except hydrogen, helium and oxygen from their solar values by the metallicity factor. On the other hand, for calculations where we define C/H by the C/O ratio we instead uniformly scale the abundances of all elements except hydrogen, helium and carbon from their solar values by the metallicity factor. Importantly we note that the total mass fraction of metals ($Z$) is not the same as the metallicity value that we quote since $Z$ will also depend on the C/O ratio, a point previously mentioned by \citet{MosLV13}.

In \cref{section:1d,section:spectra} we present results showing the atmospheric thermal and chemical structure, as well as the spectra, for cases of C/O = 0.1 and C/O = 2. For each C/O ratio case, we investigate the differences between cases where O/H is held fixed at its solar value while C/H is varied and the alternative that C/H is held fixed to its solar value while O/H is varied. For clarity, the values of C/H, O/H and N/H that used for these calculations are shown in \cref{table:ratios} for each case.

\section{Results}
\label{section:forward}

\subsection{Equilibrium composition as a function of the C/O ratio}
\label{section:box}

\begin{figure*}
  \begin{center}
  \includegraphics[width=0.49\textwidth]{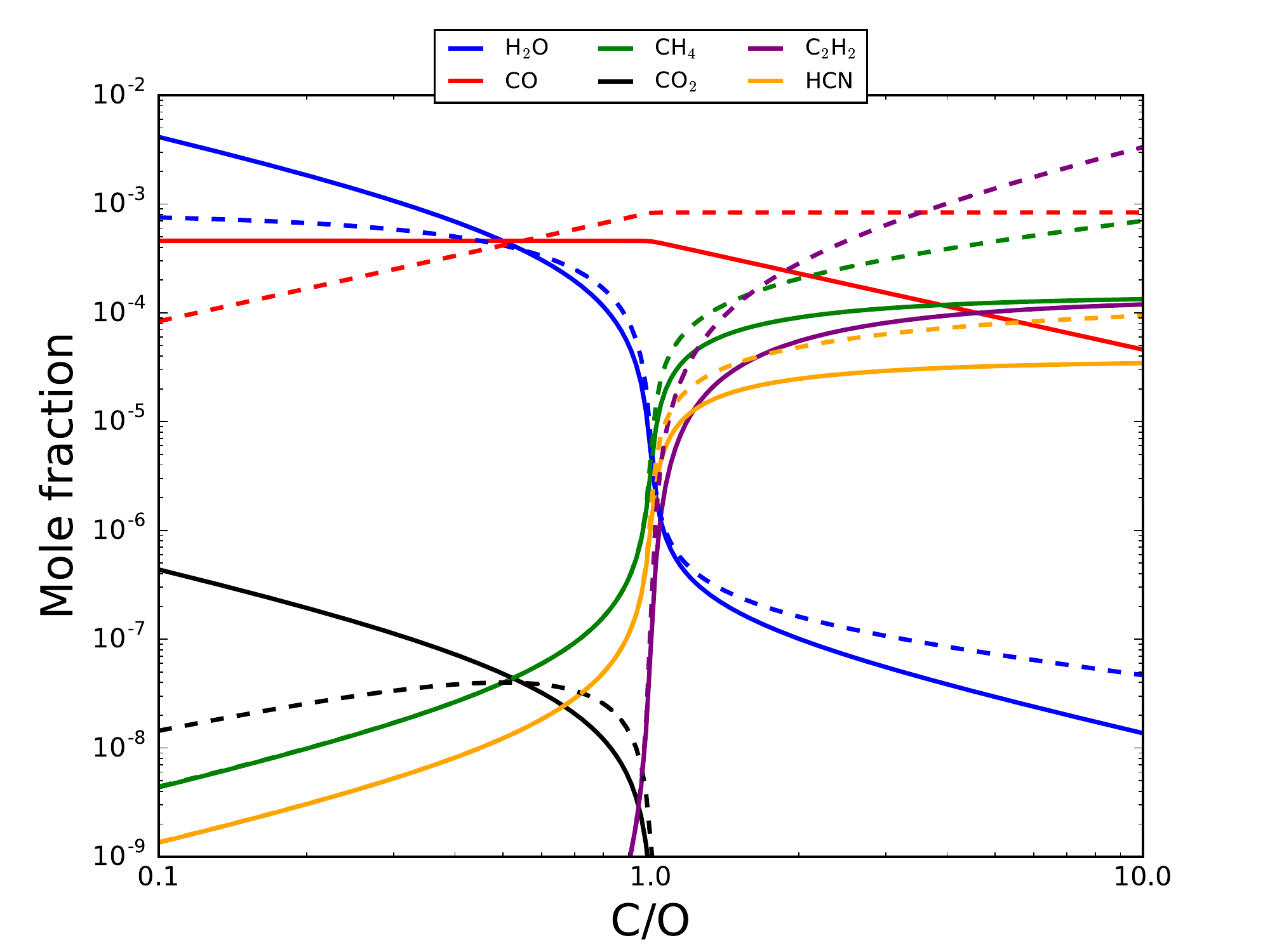} 
  \includegraphics[width=0.49\textwidth]{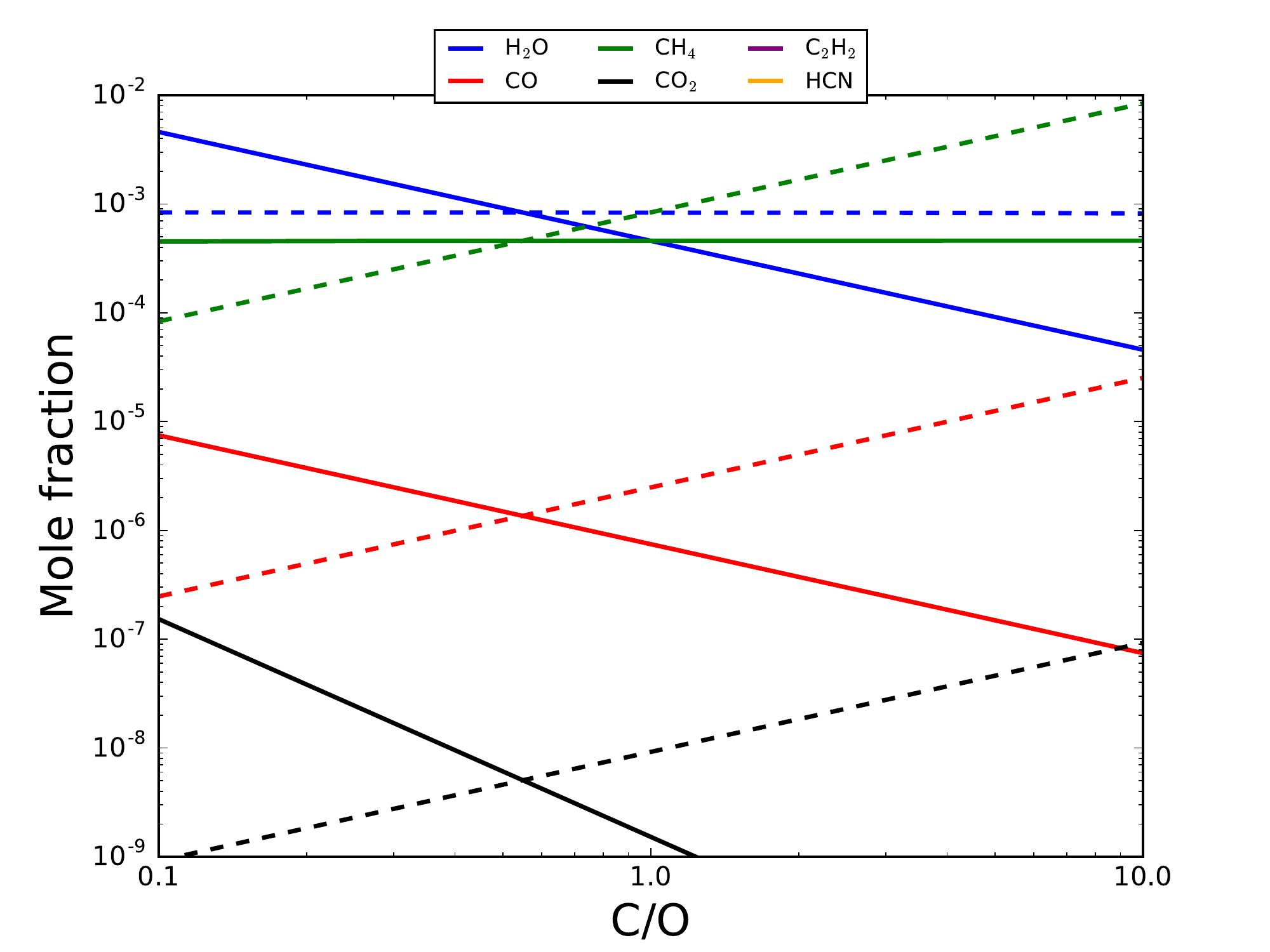} \\
  \includegraphics[width=0.49\textwidth]{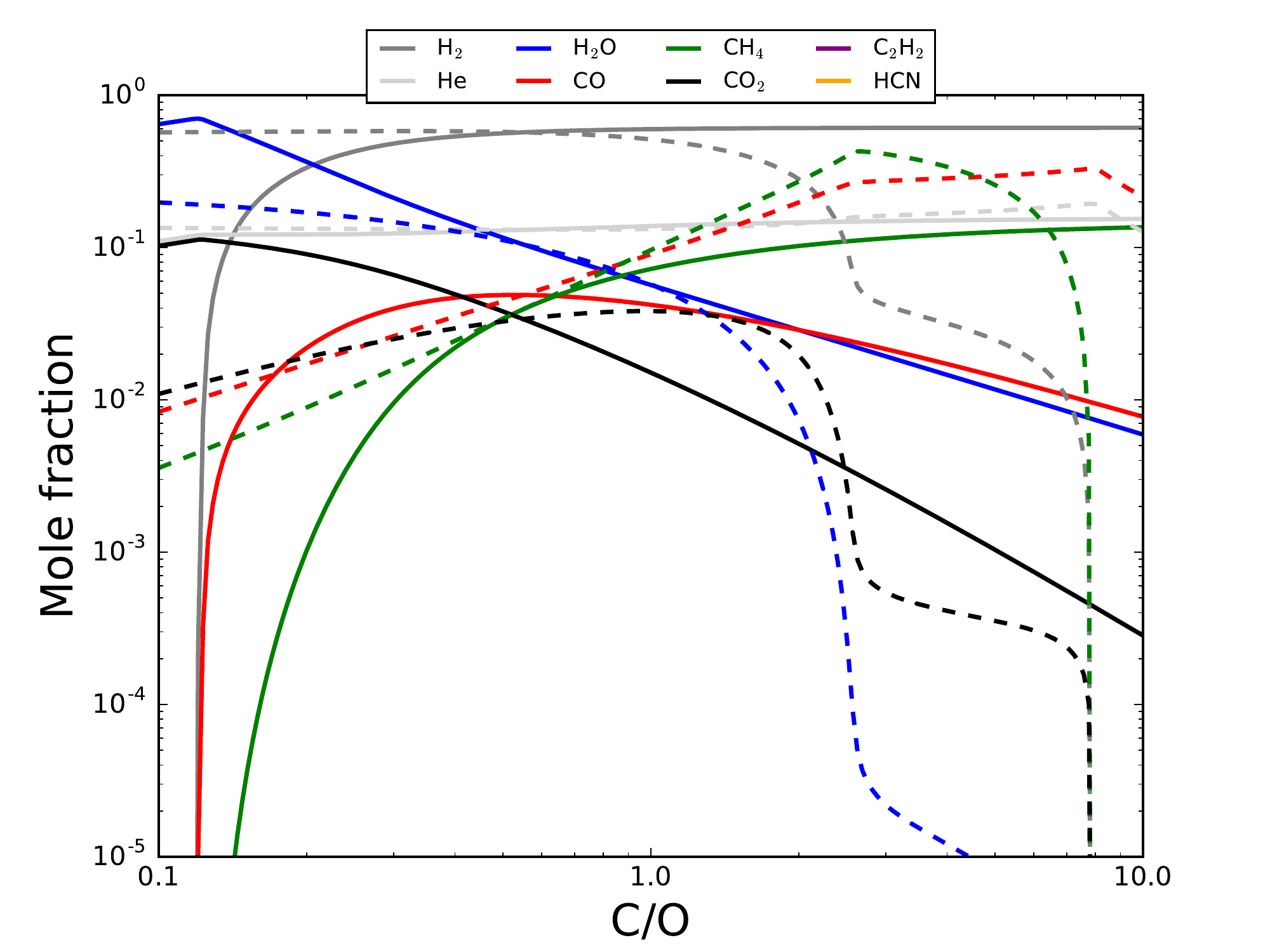} 
\includegraphics[width=0.49\textwidth]{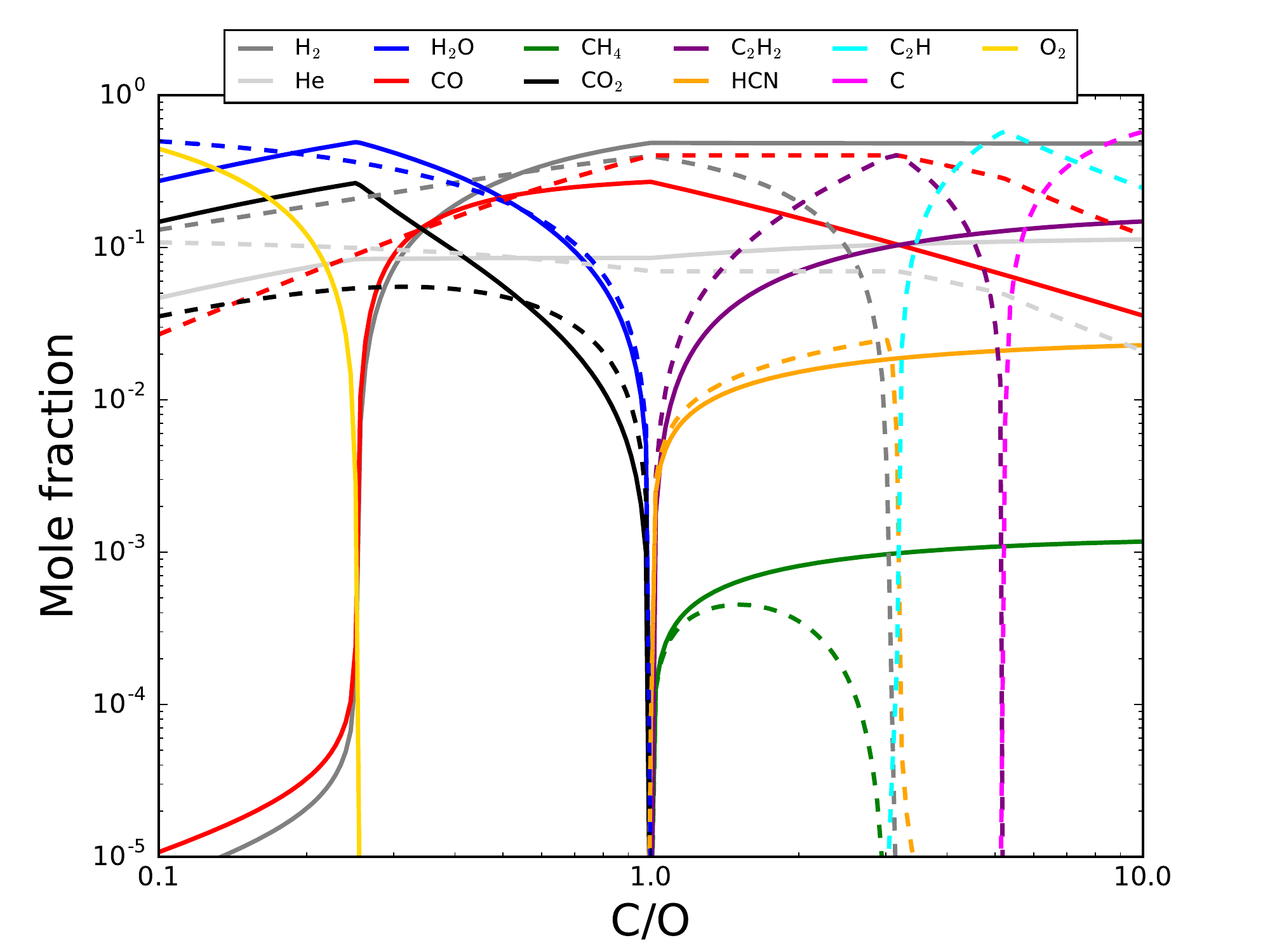} 

  \caption{Mole fractions of several key chemical species as a function of C/O ratio for different pressures, temperatures and metallicities: $P=1$ bar, $T=2000$ K and $1\times$ solar metallicity (top left), $P=0.1$ bar, $T=800$ K and $1\times$ solar metallicity (top right) and $P=0.1$ bar, $T=800$ K, $300\times$ solar metallicity (bottom left) and $P=0.01$ bar, $T=1500$ K and $1000\times$ solar metallicity (bottom right). Results obtained by varying  O/H with a fixed C/H are shown with solid lines while those obtained by varying C/H with a fixed O/H are shown in dashed lines.}
  \label{figure:box}
\end{center}
\end{figure*}

Here we consider the dependence of the chemical equilibrium composition on the C/O ratio for several combinations of pressure, temperature and metallicity. \cref{figure:box} shows the mole fractions of several key chemical species as a function of the C/O ratio for the cases where we assume a fixed O/H (the solar value shown in \cref{table:ratios}) and vary C/H and also where we assume a fixed C/H and vary O/H. 

The top left panel of \cref{figure:box} shows the chemical equilibrium mole fractions as a function of the C/O ratio for $P=1$ bar, $T=2000$ K and a solar metallicity, which match the parameters of Fig. 1 in \citet{Mad12}. Our result from the calculation that varies C/H with a fixed O/H matches well with the result of \citet{Mad12}, with a clear transition at C/O$~=~1$ where the mixture changes from being oxygen-rich (i.e. O/H$>$C/H) to carbon-rich (O/H$<$C/H). For this pressure and temperature CO is the preferred form of carbon, as opposed to CH$_4$ \citep[e.g.][]{FegL96}. Where C/O$~<1$, oxygen atoms in the mixture that are not locked up in CO result in a large abundance of water (H$_2$O). However, for C/O$~>1$ most of the oxygen atoms are locked into forming CO and the abundance of H$_2$O is much lower. The remaining carbon atoms that are not locked up in CO are primarily in the form of CH$_4$, acetylene (C$_2$H$_2$) and hydrogen cynanide (HCN), which all have much larger abundances compared with C/O$~<1$.

We find both quantitative and qualitative differences in the chemical equilibrium mole fractions between the cases where we either vary O/H or vary C/H to alter the C/O ratio. For C/O $<1$ the CO mole fraction is limited by the amount of carbon, since carbon is less abundant than oxygen. Therefore, for the calculation where C/H is fixed, and the C/O ratio is altered by varying O/H, the mole fraction of CO is independent of the C/O ratio. On the other hand, for the calculation where O/H is fixed and C/H is varied the mole fraction of CO increases with increasing C/O ratio. For the same reason, carbon dioxide (CO$_2$) becomes more abundant with decreasing C/O ratio when O/H is increased, since more oxygen is available, but when C/H is decreased CO$_2$ becomes less abundant with decreasing C/O ratio, since less carbon is available. Additionally, as O/H is increased to achieve a smaller C/O ratio the extra oxygen atoms lead to a larger H$_2$O abundance compared with the calculation where O/H is assumed to be fixed.

For C/O $>1$ the mole fraction of CO is instead limited by the amount of oxygen, which is now less abundant than carbon. Therefore, as the C/O ratio is increased by decreasing O/H (with a fixed C/H) the mole fraction of CO decreases. However, for the calculation where the C/O ratio is increased by increasing C/H (with a fixed O/H) the mole fraction of CO is independent of the C/O ratio. The mole fractions of the carbon-bearing species CH$_4$, C$_2$H$_2$ and HCN are all larger in the calculation where C/H is varied compared with the calculation where C/H is held fixed, since the number of available carbon atoms increases with increasing C/O ratio.

The top right panel of \cref{figure:box} shows the case of $P = 0.1$ bar, $T=800$ K and solar metallicity, where CH$_4$ is now the most favourable form of carbon, instead of CO, as is the case for Jupiter in our own solar system \citep[e.g.][]{Lew69}. The trends in the mole fractions with C/O ratio are much simpler in this case. For the calculation where C/H is held fixed and the C/O ratio is altered by varying O/H, the CH$_4$ mole fraction is independent of C/O ratio. On the other hand, the mole fractions of the oxygen-bearing species H$_2$O, CO and CO$_2$ decrease with increasing C/O ratio as the number of available oxygen atoms decreases. On the other hand, for the calculation where O/H is held fixed and the C/O ratio is altered by varying C/H, the H$_2$O mole fraction is independent of C/O ratio while the abundances of CH$_4$, CO and CO$_2$ increase with increasing C/O ratio as the number of carbon atoms increases.

The bottom left panel of \cref{figure:box} shows the case of $P = 0.1$ bar, $T=800$ K and 300$\times$ solar metallicity. These parameters match those of \citet[][Fig. 4, bottom row]{MosLV13}, who assume a fixed C/H and vary O/H. We reiterate from \cref{section:ele} that, for the calculation where we assume a fixed C/H when varying the C/O ratio, we have increased the abundances of all elements except H, He and O by a factor of 300, in this case. The value for O/H is then calculated using both C/H and the C/O ratio. Likewise, for the calculation where we assume a fixed O/H when varying the C/O ratio, we have increased the abundances of all elements except H, He and C by a factor of 300, with C/H then calculated using O/H and the C/O ratio.

In this case of a higher (300$\times$ solar) metallicity there is a more pronounced difference in the mole fractions of many chemical species, at a given C/O ratio, depending on whether we vary O/H or C/H. When varying O/H, for very small values of the C/O ratio oxygen becomes almost as abundant as hydrogen; at C/O = 0.1 we have O/H $\sim0.8$. Here the oxygen-bearing species H$_2$O, CO and molecular oxygen (O$_2$) become the most abundant chemical species in the mixture with a very low abundance of molecular hydrogen (H$_2$), in good agreement with the results of \citet{MosLV13}. On the other hand, if we instead alter the C/O ratio by varying C/H then H$_2$ is instead the most abundant species at C/O = 0.1. Though the C/O ratio is the same the oxygen abundance is significantly smaller at O/H $\sim0.1$, compared with the previous case. For high values of the C/O ratio carbon similarly becomes almost as abundant as hydrogen; in fact, at C/O = 10 we have C/H $\sim1.5$ with C$_2$H and CO as the most abundant chemical species. In contrast, when varying O/H with a fixed C/H, H$_2$ and CH$_4$ are the most abundant species at C/O = 10.

These results support the previous findings of \citet{MosLV13,HenL16,HenT16} and show that the chemical equilibrium composition can be significantly different for the same C/O ratio but with different carbon and oxygen abundances relative to hydrogen.

\subsection{Mean molecular mass as a function of C/O}
\label{section:mmm}

\begin{figure}
  \center
   \includegraphics[width=0.49\textwidth]{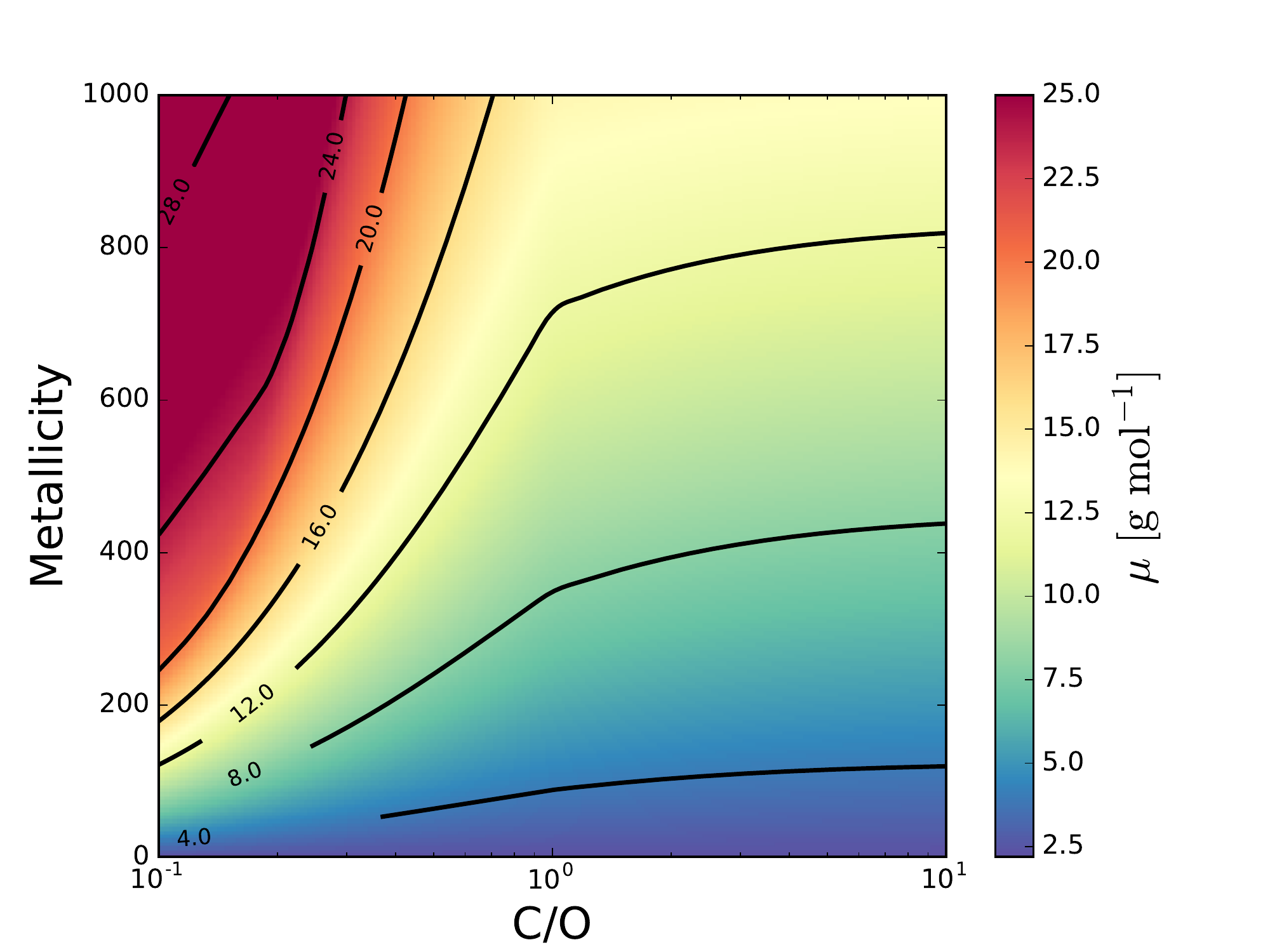} \\
  \includegraphics[width=0.49\textwidth]{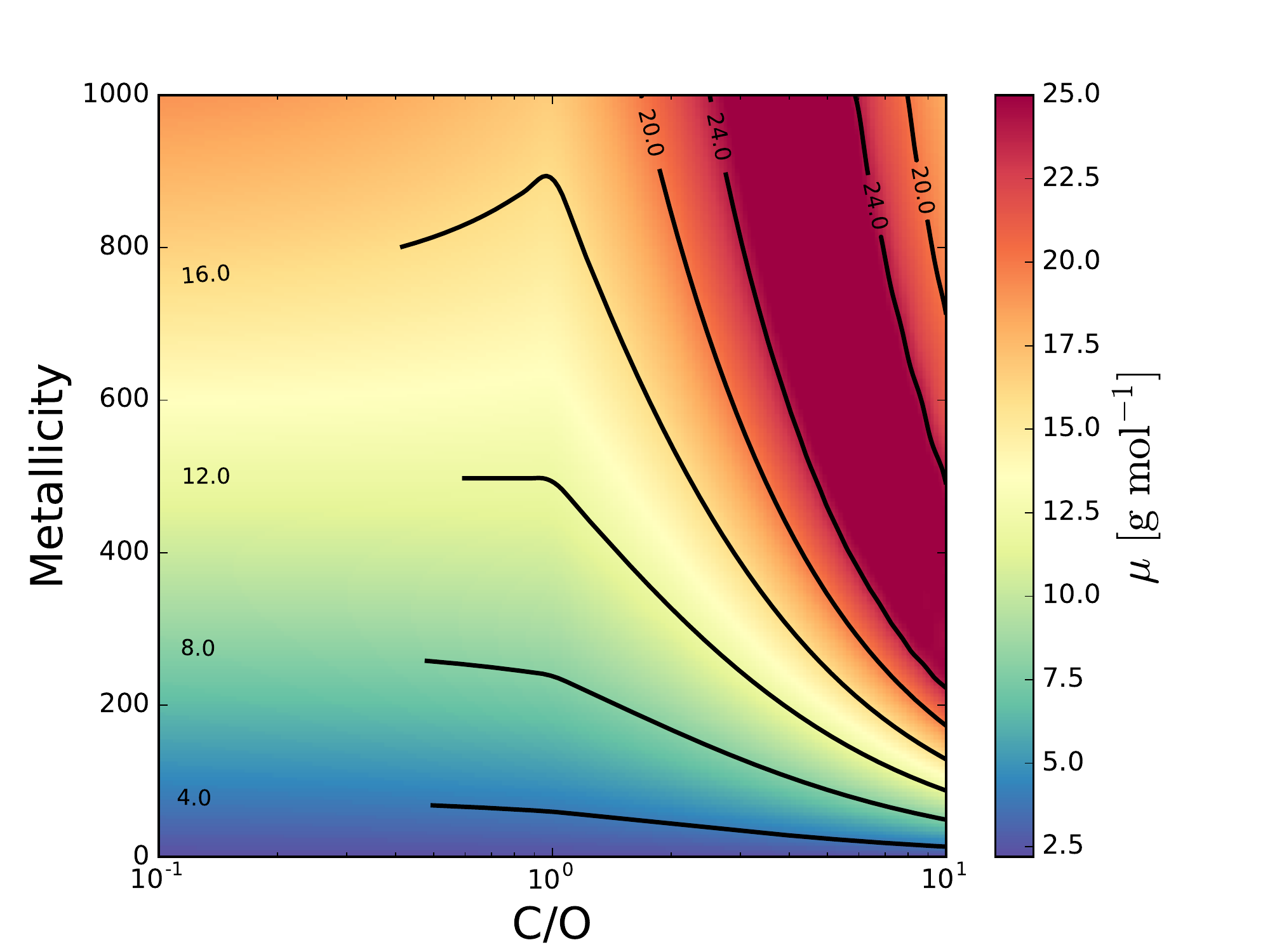} \\
  \caption{Mean molecular mass (colour scale and black contours) as a function of metallicity and C/O ratio for a calculation that adjusts the C/O ratio by varying O/H with the value of C/H determined by the metallicity (top) and for a calculation that adjusts the C/O ratio by varying C/H with the value of O/H determined by the metallicty (bottom), both with $P=10$ mbar and $T=1500$ K.}
  \label{figure:mu}
\end{figure}

The mean molecular mass ($\mu$) of the gas is determined by the chemical composition. \cref{figure:mu} shows $\mu$ as a function of metallicity and the C/O ratio for $P=10$ mbar and $T=1500$ K. In one case we assume that the C/H is defined by the metallicity and its solar element abundance and O/H is then calculated from C/H and the C/O ratio. In the other case we assume that O/H is instead defined by the metallicity and its solar element abundance while C/H is calculated from O/H and the C/O ratio. The difference in $\mu$ between these two cases can be as large as a factor of two, with the difference being largest for small (${\leq0.3}$) and large (${\geq2}$) values of the C/O ratio. However, the difference can still be significant for C/O ratios closer to the solar value. For example, at a metallicity of 1000$\times$ solar and C/O = 0.4, the calculation assuming C/H is tied to the metallicity results in $\mu\sim20.3$ g mol$^{-1}$ whereas the calculation assuming that O/H is tied to the metallicity results in $\mu\sim18.1$ g mol$^{-1}$.

For the calculation where O/H is defined by the C/O ratio, $\mu$ increases with decreasing C/O ratio. As O/H increases toward smaller values of the C/O ratio, eventually oxygen becomes the most abundant element in the mixture and the dominant chemical species is O$_2$ followed by H$_2$O, as shown in the bottom right panel of \cref{figure:box}. On the other hand, for the calculation where C/H is defined by the C/O ratio $\mu$ increases with increasing C/O ratio; though it begins to decrease again for very high C/O ratios. Carbon becomes the most abundant element for large C/O and the most abundant chemical species for C/O $\sim6$ is C$_2$H followed by CO. For larger C/O ratios still, atomic carbon (C) becomes the most abundant chemical species, which is why $\mu$ decreases for C/O $>0.8$.

\subsection{1D atmospheric simulations: composition and thermal structure}
\label{section:1d}

In this section we investigate the effect of assuming different sets of element abundances that have the same C/O ratio with 1D atmospheric simulations. We consider two cases: a ``hot'' atmosphere, where CO is the most stable carbon species in chemical equilibrium, and a ``warm'' atmosphere, where CH$_4$ is the most stable carbon species in chemical equilibrium. Throughout this section we will refer to the calculations that assume the solar element abundances as the ``solar case''. We also investigate cases of C/O = 0.1 and C/O = 2 by adjusting the set of element abundances assuming either a solar O/H or a solar C/H. Abundance ratios of carbon, oxygen and nitrogen for these different sets of element abundances are shown in \cref{table:ratios}.

\subsubsection{Hot atmosphere}

\begin{figure}
  \center
  \includegraphics[width=0.5\textwidth]{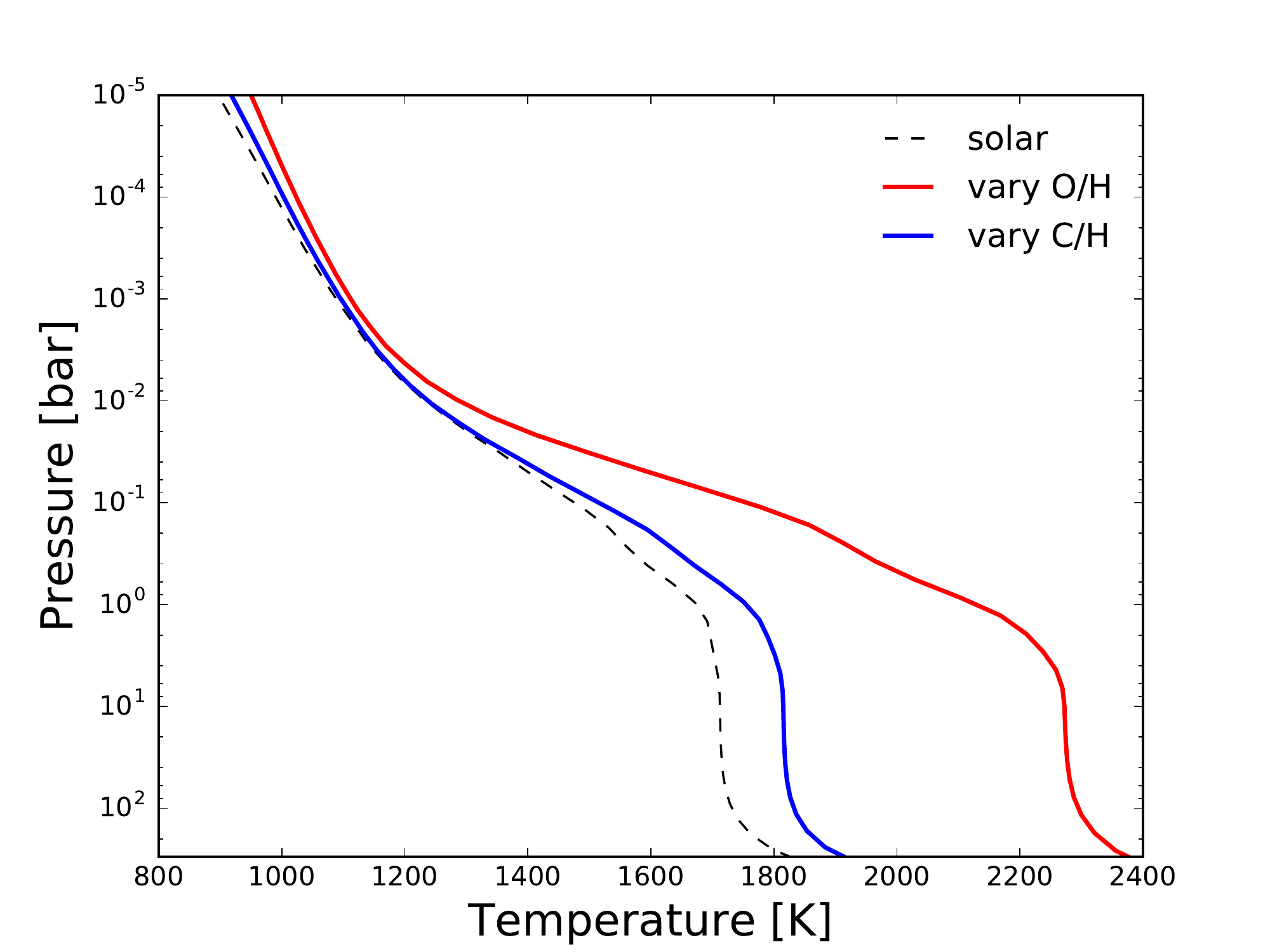} \\
\includegraphics[width=0.5\textwidth]{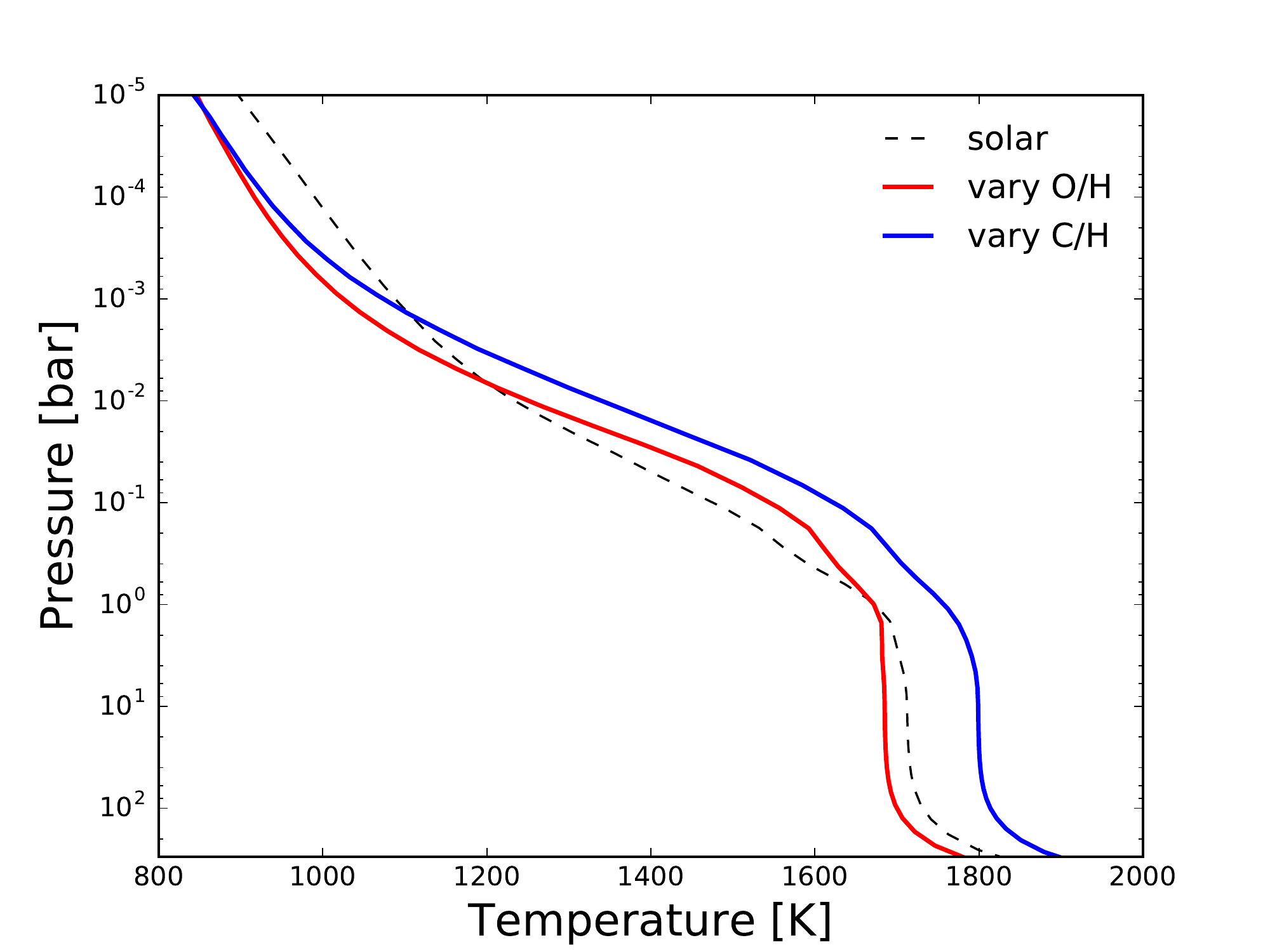}
  \caption{$P$-$T$ profiles for the hot atmosphere for C/O = 0.1 (top) and C/O = 2 (bottom). The $P$-$T$ profile assuming solar element abundances is also shown in both panels.}
  \label{figure:ptwarm}
\end{figure}

\cref{figure:ptwarm} shows the $P$-$T$ profiles for the hot atmosphere for different sets of element abundances. For both C/O = 0.1 and C/O = 2, the $P$-$T$ profile is significantly different between the calculations that assume solar O/H or solar C/H. For C/O = 0.1, the atmosphere is more than 400 K warmer for $P>1$ bar when increasing O/H, compared with when decreasing C/H. For lower pressures ($P<0.1$ bar), where transmission spectroscopy probes, the temperature difference can be $\sim100$ K. For C/O = 2, there are also temperature differences of more than 100 K over a very large pressure range. Notably, for pressures larger than 1 bar, the temperature is cooler than the solar profile when decreasing O/H while when increasing C/H the temperature is warmer than the solar profile. To understand the changes in the temperature between each of these cases we must first consider the chemical composition.

\begin{figure}
  \center
  \includegraphics[width=0.5\textwidth]{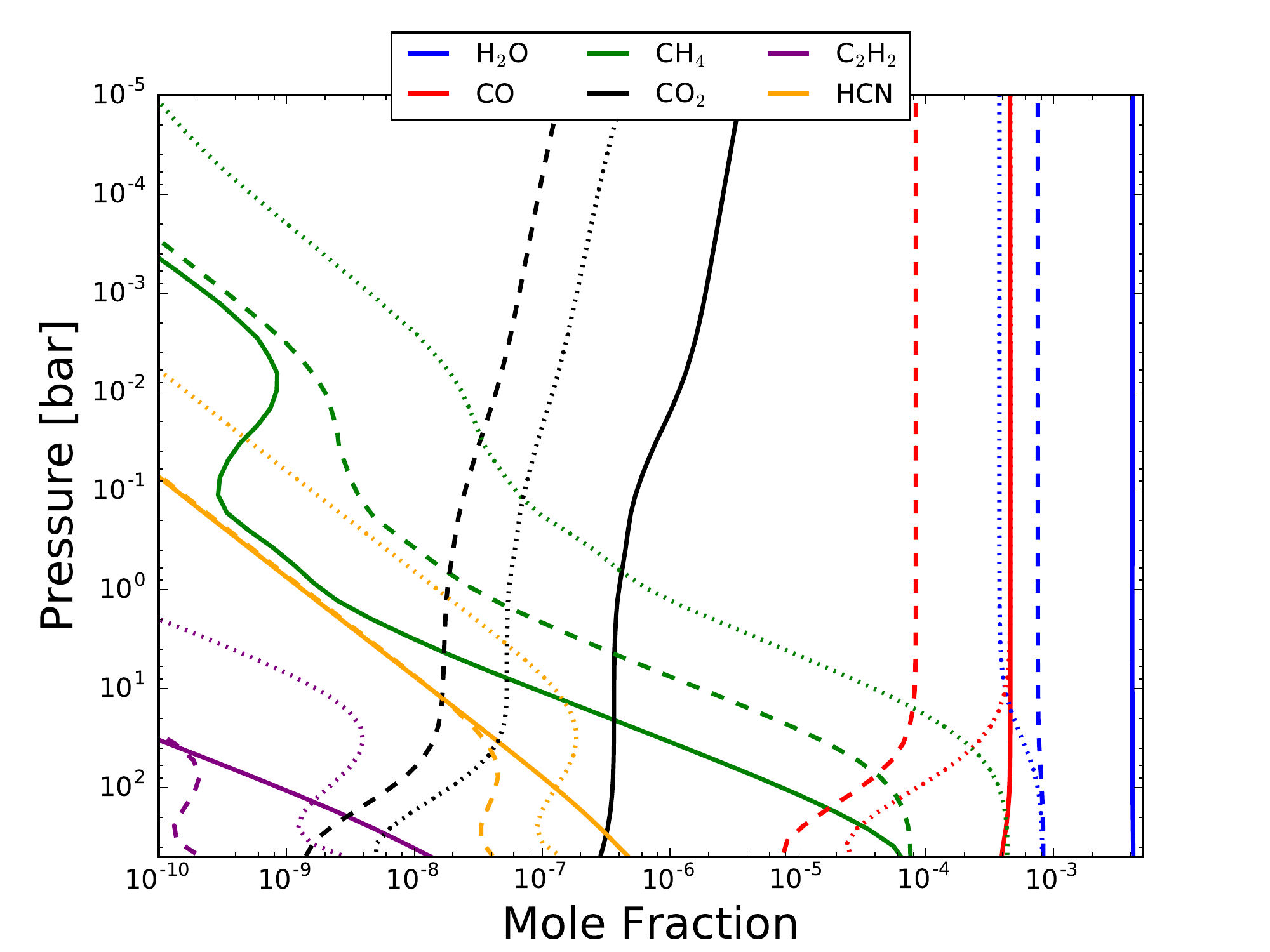}\\
  \includegraphics[width=0.5\textwidth]{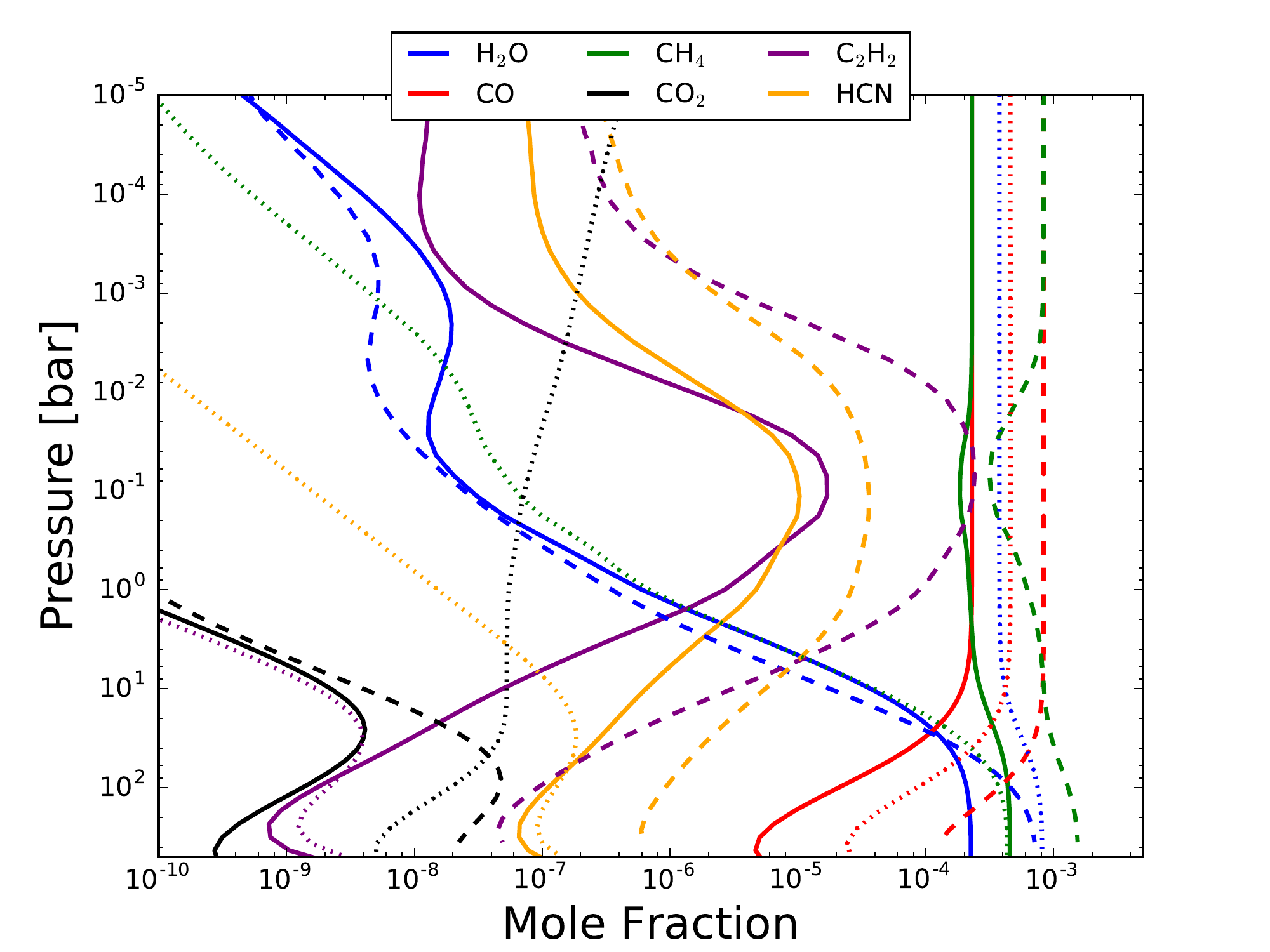} 
  \caption{Mole fractions of key chemical species for the hot atmosphere with C/O = 0.1 (top) and C/O = 2 (bottom) for the model that varies O/H (solid) and the model that varies C/H (dashed). The mole fractions assuming solar element abundances (dotted) are also shown in both panels.}
  \label{figure:chemwarm}
\end{figure}

\cref{figure:chemwarm} shows the chemical equilibrium mole fractions of several key species for the hot atmosphere. For C/O = 0.1, compared with the solar case, the abundance of H$_2$O increases both when increasing O/H with a fixed C/H and when decreasing C/H with a fixed O/H, though the increase is much larger when increasing O/H than when decreasing C/H. When increasing O/H the H$_2$O abundance increases simply because O/H is increased and more oxygen atoms are available to form H$_2$O. On the other hand, when decreasing C/H the H$_2$O abundance increases because less CO can form, due to there being fewer carbon atoms, meaning more oxygen atoms are available to form H$_2$O. The mole fraction of CO is equivalent to the solar case (over most of the modelled pressure range) when increasing O/H because its abundance is limited by the amount of carbon and C/H is unchanged. For the same reason, when decreasing C/H the mole fraction of CO is significantly lower than for the solar case.
 
The CH$_4$ mole fraction is reduced compared with the solar case for both cases of increasing O/H and decreasing C/H, though with a larger reduction when increasing O/H. We have already shown in \cref{figure:box} (top panel) that for C/O $<1$ and for high temperatures (where CO is favoured over CH$_4$) that the CH$_4$ mole fraction is not sensitive to the C/O ratio. The difference in the CH$_4$ abundance shown in \cref{figure:chemwarm} is a temperature effect, with a warmer temperature resulting in a smaller equilibrium CH$_4$ mole fraction. The CO$_2$ mole fraction is greater than for the solar case when increasing O/H, due to a greater availability of oxygen atoms, but is smaller than for the solar case when decreasing C/H, due to there being fewer carbon atoms. 

The change in temperature already shown in \cref{figure:ptwarm} is largely explained by the change in the abundance of H$_2$O, which has strong absorption bands across a broad spectral range. When decreasing the C/O ratio to 0.1, either by increasing O/H or decreasing C/H, a larger H$_2$O mole fraction increases the opacity. However, the calculation that increases O/H results in a greater increase in the H$_2$O abundance, explaining the larger temperature increase.

\cref{figure:chemwarm} also shows the mole fractions of the most abundant chemical species for the hot atmosphere with C/O = 2. The CH$_4$ abundance increases significantly compared with the solar case, for both calculations that decrease O/H and increase C/H, while the H$_2$O abundance shows a similarly significant decrease. CO is the favoured form of carbon, over CH$_4$, for all but the highest pressures in this model. For C/O$~<1$, the abundance of CO is limited by the carbon abundance, while for C/O$~>1$ it is limited by the oxygen abundance. Decreasing O/H therefore leads to a smaller CO mole fraction, compared with the solar case, while increasing C/H leads to a larger CO mole fraction, as previously shown in \cref{figure:box} (top panel). Compared with the solar case the abundances of acetylene and hydrogen cyanide are also larger for C/O = 2. However, the mole fractions of each of these species are $\sim1$ order of magnitude larger for the calculation where C/H is increased compared with the calculation where O/H is decreased, due to a larger availability of carbon atoms.

The difference in the temperature profiles between the models that vary O/H or vary C/H are explained by the difference in the CH$_4$ abundance. Compared with the solar case CH$_4$ has replaced H$_2$O as the dominant absorber in the atmosphere. The CH$_4$ mole fraction is larger for the calculation that increases C/H, compared with the calculation that decreases O/H, leading to a larger atmospheric opacity and higher temperatures in the former.

We note that the abundances presented \cref{figure:chemwarm} are calculated consistently with the $P$-$T$ profiles shown in \cref{figure:ptwarm}. Therefore changes to the chemical equilibrium mole fractions compared with the solar case may be due to a combination of the altered element abundances as well as changes in the $P$-$T$ profile.

\subsubsection{Warm atmosphere}

\begin{figure}
  \center
  \includegraphics[width=0.5\textwidth]{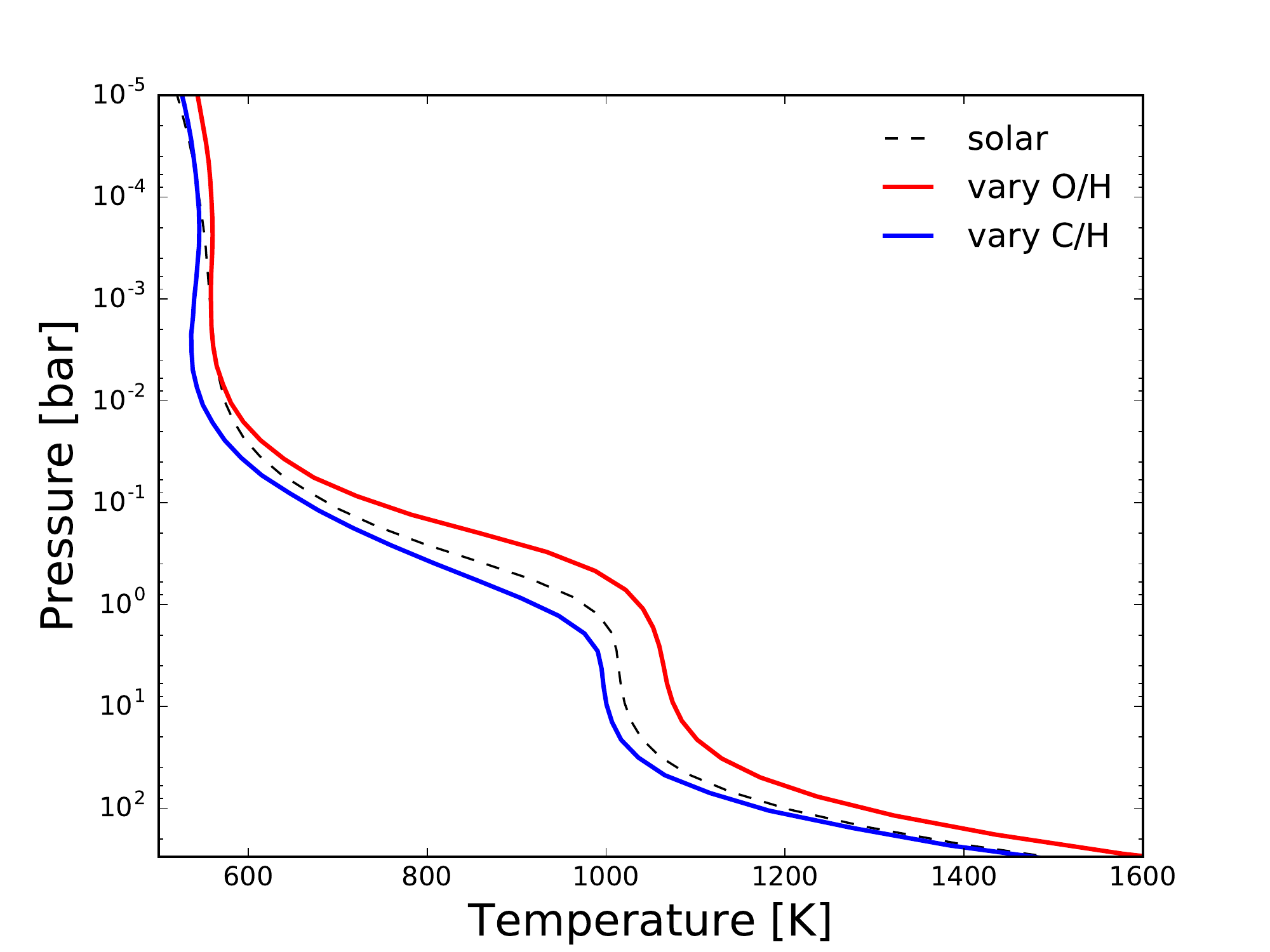} \\
  \includegraphics[width=0.5\textwidth]{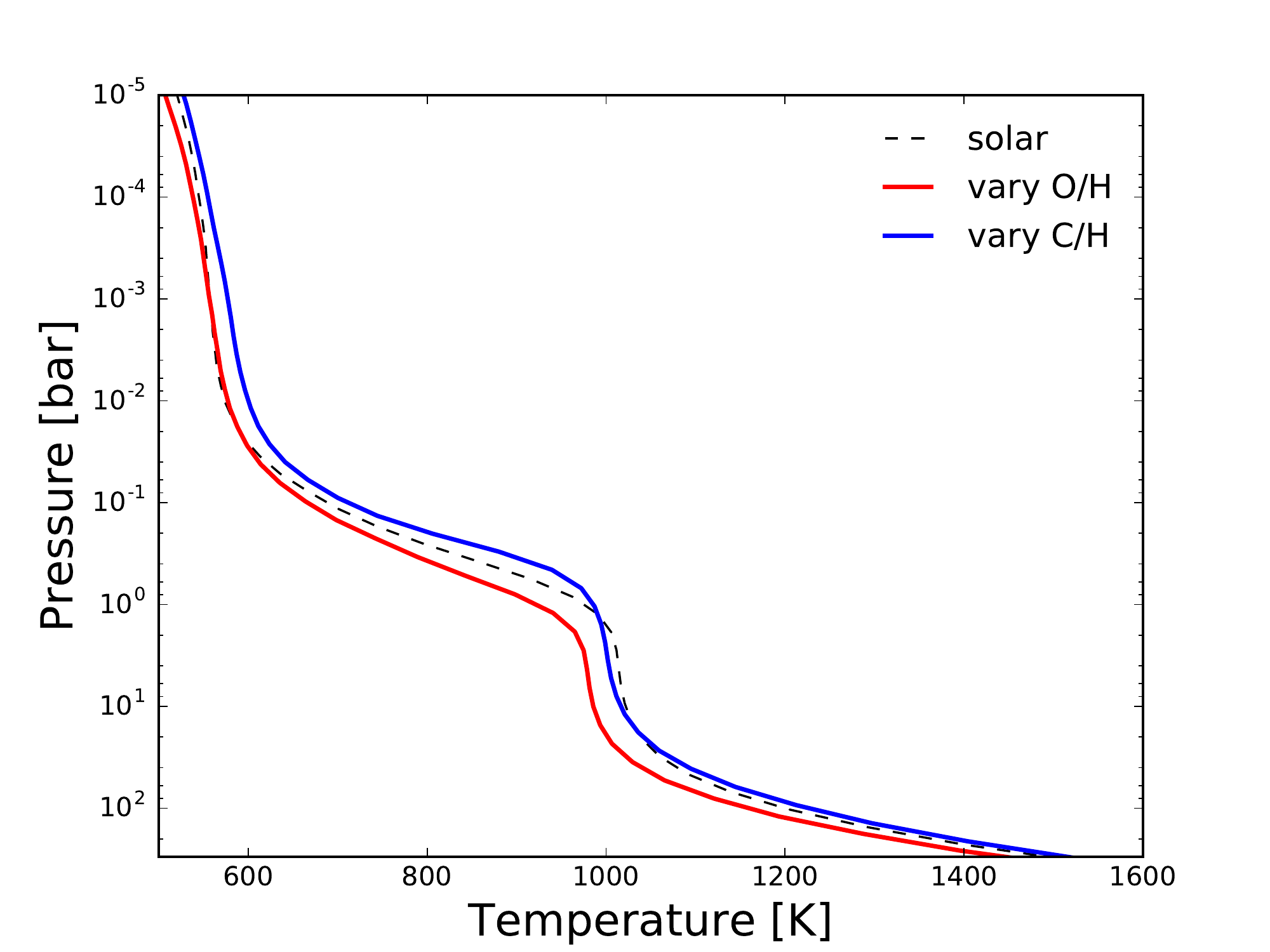}
  \caption{As \cref{figure:ptwarm} but for the warm atmosphere}
  \label{figure:ptcool}
\end{figure}

\cref{figure:ptcool} shows the $P$-$T$ profiles for the warm atmosphere. Differences between the calculations that vary O/H and that vary C/H are significant, with temperature differences of up to 100 K for both C/O = 0.1 and C/O = 2. Notably for C/O = 0.1 the temperature profile is hotter than for the solar case when increasing O/H but is cooler than the solar case when decreasing C/H.

\begin{figure}
  \center
  \includegraphics[width=0.5\textwidth]{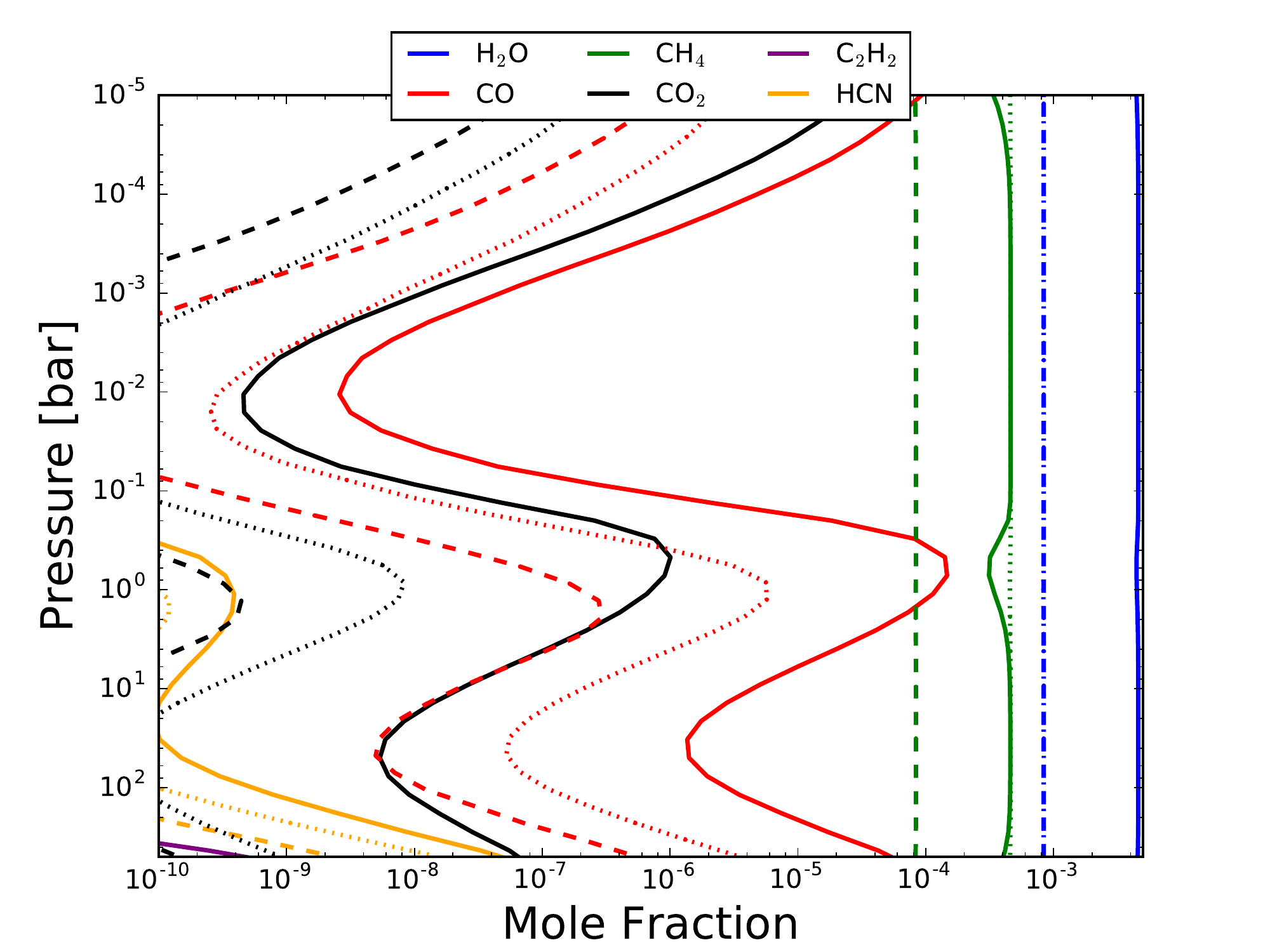} \\
  \includegraphics[width=0.5\textwidth]{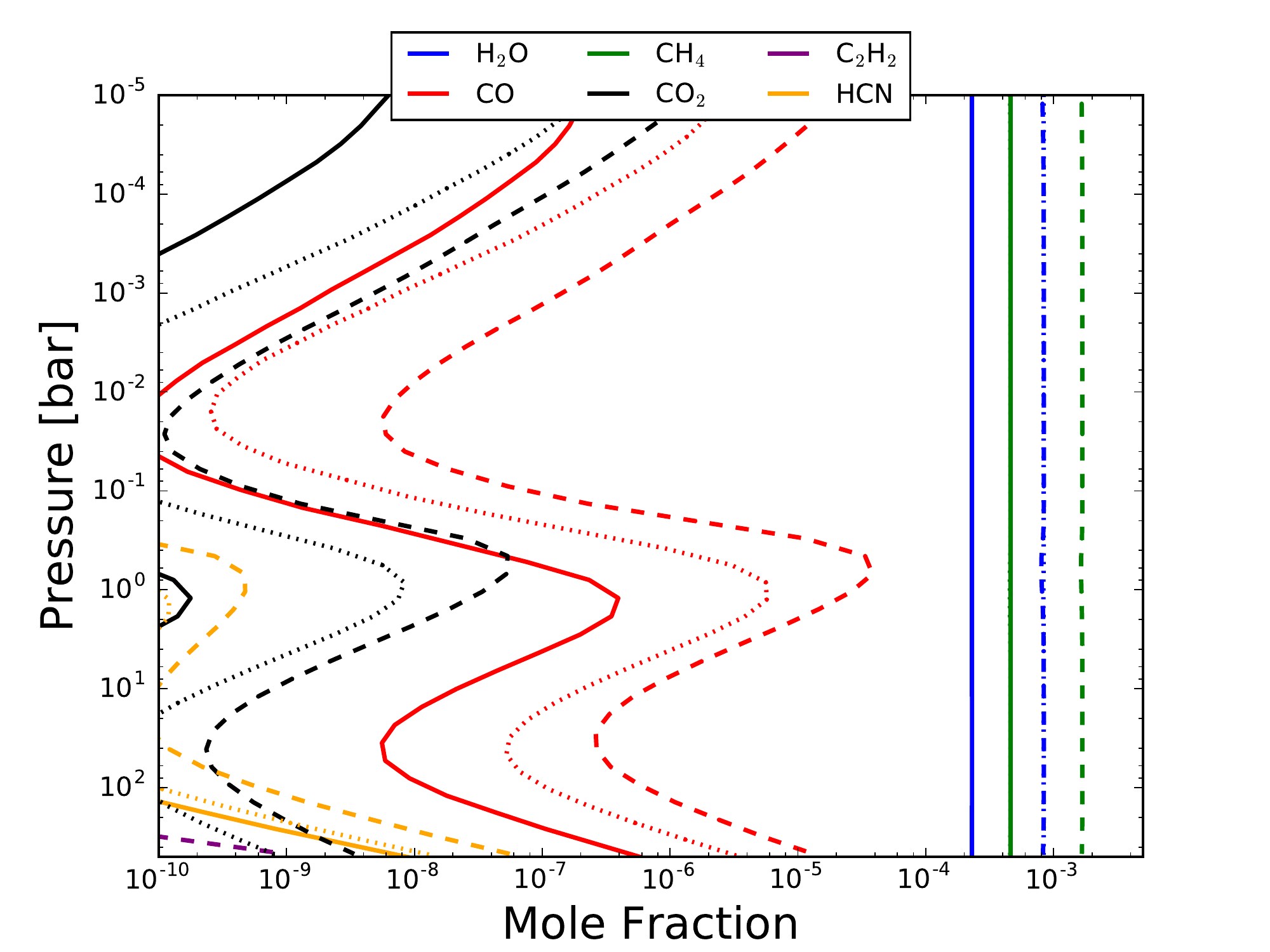}
  \caption{As \cref{figure:chemwarm} but for the warm atmosphere.}
  \label{figure:chemcool}
\end{figure}

\cref{figure:chemcool} shows the equilibrium chemical composition for each of the $P$-$T$ profiles shown in \cref{figure:ptcool}. As with the hot atmosphere, there are significant differences in the mole fractions between the calculations that vary O/H and that vary C/H. An important difference with the hot atmosphere presented in the previous section is that CH$_4$ is now the most favourable form of carbon, instead of CO \citep{Lew69}.

For the C/O = 0.1 case, the H$_2$O abundance increases significantly when increasing O/H, due to a larger abundance of oxygen, while the CH$_4$ abundance is almost unchanged, since the amount of carbon has not changed. This increase in the H$_2$O mole fraction increases the atmospheric opacity, leading to the hotter temperature profile shown in \cref{figure:ptcool}. On the other hand, for the calculation that decreases C/H the H$_2$O abundance is approximately the same as for the solar case while the CH$_4$ mole fraction has reduced, due to fewer carbon atoms being available. This reduction in the CH$_4$ abundance decreases the atmospheric opacity and leads to the cooler temperature profile, compared with the solar case, as shown in \cref{figure:ptcool}. The abundances of CO and CO$_2$ increase for the calculation that increases O/H but decrease for the calculation that decreases C/H and may also contribute to the differences in the temperature structure between the two cases.

There are similar trends apparent for the C/O = 2 case, though in the opposite sense. For the calculation that decreases O/H the H$_2$O abundance is less than for the solar case, due to there being less availability of oxygen atoms, while it is approximately unchanged for the calculation that increases C/H with a fixed O/H. Likewise, the abundance of CH$_4$ is largely unaffected while C/H is held fixed but increases significantly when C/H is increased. The abundances of CO and CO$_2$ decrease when O/H is decreased with a fixed C/H, due to there being less oxygen available, but increase when C/H is increased with a fixed O/H, due to there being more carbon atoms available.

\subsection{1D atmospheric simulations: emission and transmission spectra}
\label{section:spectra}

\begin{figure*}
  \center
  \includegraphics[width=\textwidth]{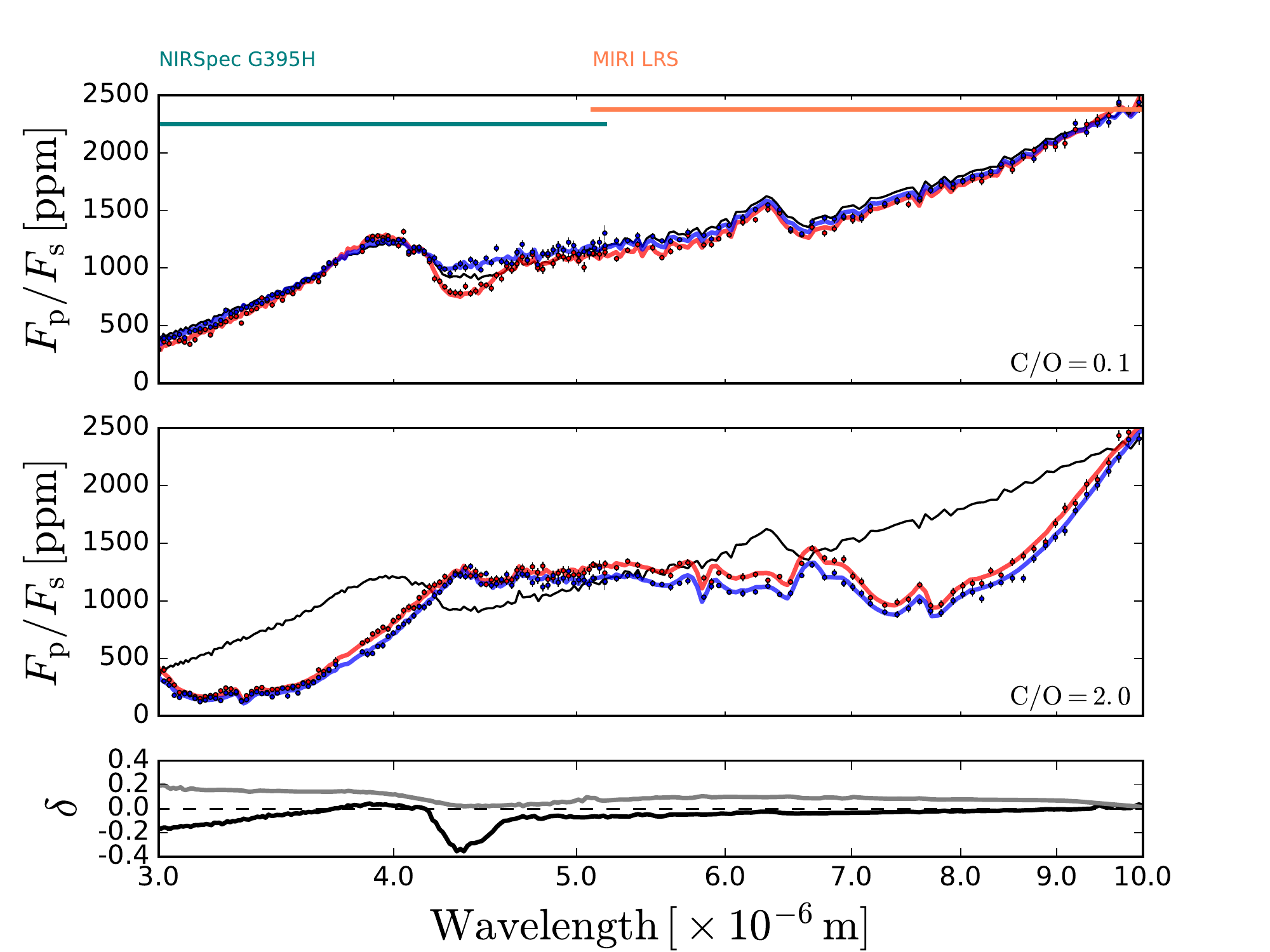}
   \caption{Emission spectra for the hot atmosphere for C/O = 0.1 (top) and C/O  = 2 (middle) for the model that varies O/H (red) and the model that varies C/H (blue). The emission spectrum for the solar element abundances simulation (black) is also shown in both panels (top and middle). Also shown are PandExo simulated observations for the NIRSpec G395H and MIRI LRS modes ({\it JWST}), with the wavelength coverage indicated by the coloured horizontal lines in the top panel. The bottom panel shows the fractional difference in the spectra between the models that vary O/H and the models that vary C/H for C/O = 0.1 (black) and C/O = 2 (grey). The black dashed line indicates a relative difference of $\delta = 0$.}
  \label{figure:emisswarm}
\end{figure*}

In this section we present emission and transmission spectra calculated using the 1D atmospheric profiles shown in the previous section. We include simulated {\it Hubble Space Telescope} ({\it HST}) and {\it James Webb Space Telescope} ({\it JWST}) observations, calculated using the PandExo tool \citep{BatMP17}. 

The PandExo simulations were performed for the {\it JWST} NIRSpec G395H and MIRI LRS modes for emission spectra and for the {\it JWST} NIRSpec G140H, G235H and G395H and MIRI LRS modes, as well as the {\it HST} Wide Field Camera 3 (WFC3) G102 and G141 modes, for transmission spectra. With the exception of the warm atmosphere emission spectra, a single eclipse with equal in to out of transit observation time was used, with a noise floor of 50 ppm and a detector saturation set to 80\% full well. For the warm atmosphere emission spectra two transits were used. For the PandExo simulations of the {\it HST} WFC3 instrument, all necessary orbital and exposure parameters for GJ~436b were taken from \citet{KnuBD14} and for HD~209458b from \citet{Deming2013}. Stellar and planetary parameters for HD~209458b and GJ~436b were taken from the TEPCAT database\footnote{\url{http://www.astro.keele.ac.uk/jkt/tepcat/}} for the hot and warm atmosphere cases, respectively. All instrument-related parameters were kept at the PandExo defaults. 

We reiterate that the calculations presented in this paper are only loosely based on HD~209458b and GJ~436b, for the hot and warm cases, respectively. Therefore, these spectra are not intended as our best attempt at predictions of the spectra for these two specific planets. Our intention is to study the theoretical impact of varying the C/O ratio on the spectra for HD~209458b-like and GJ~436b-like planets.

\subsubsection{Hot atmosphere}

\begin{figure*}
  \center
  \includegraphics[width=\textwidth]{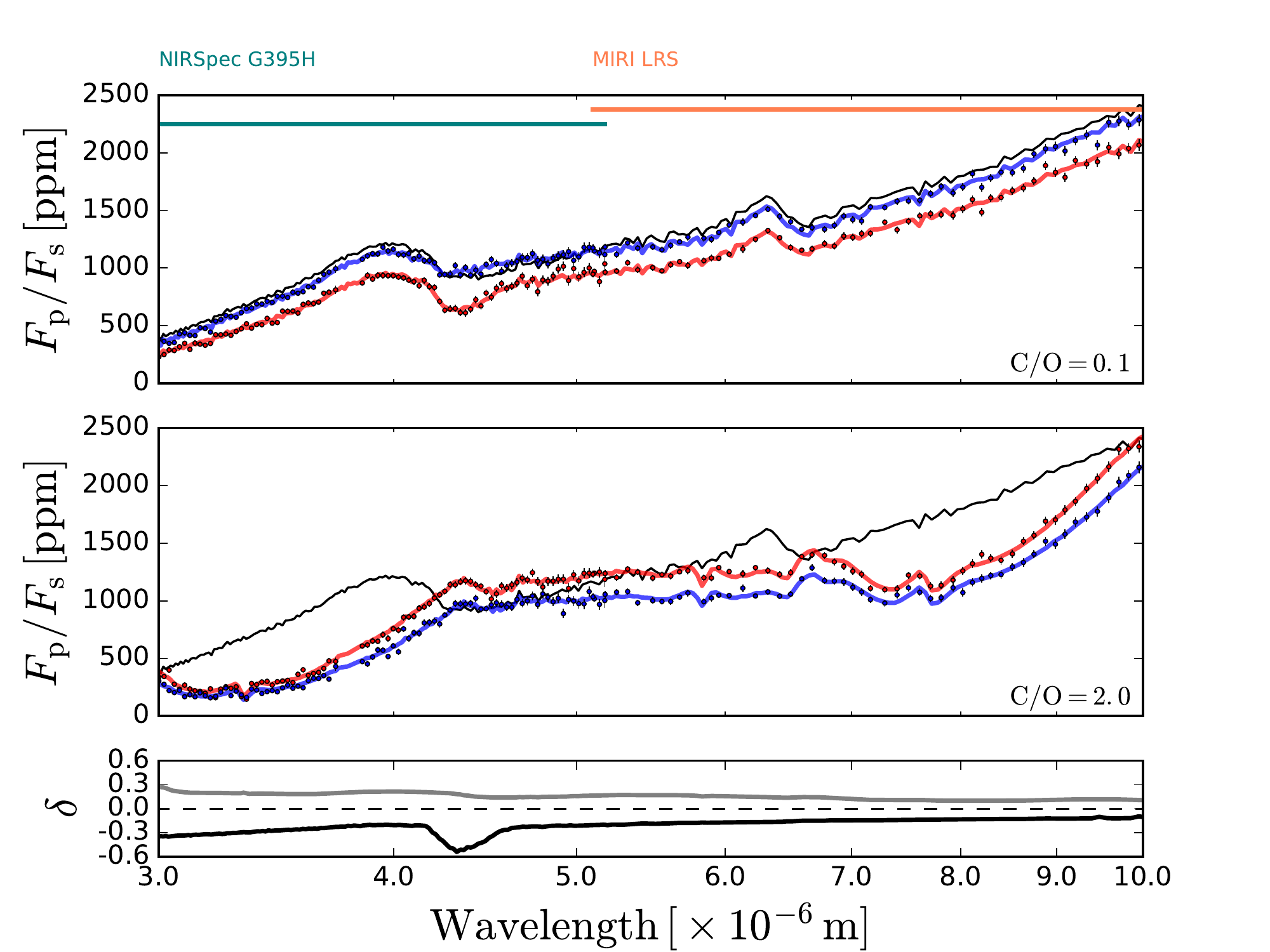}
   \caption{As \cref{figure:emisswarm} but with the chemical composition and spectrum calculated using the $P$-$T$ profile calculated using solar element abundances: the non-consistent case.}
  \label{figure:emisswarm2}
\end{figure*}

 \cref{figure:emisswarm} shows the emission spectra for the hot atmosphere. For C/O = 0.1, there is generally little difference in the spectra between the models that vary O/H or C/H, except for the region around $\sim4.3~\mu$m. Here there is a significant qualitative and quantitative difference, as the calculation that increases O/H gives a larger flux ratio compared with the solar case while the calculation that decreases C/H gives a smaller flux ratio compared with the solar case. The relative difference (defined as $\delta = (A^{\rm vary O/H} - A^{\rm  vary C/H})/A^{\rm vary O/H}$, where $A$ is the flux ratio in this case) in the flux ratio between the two cases is more than 30\%. The PandExo simulated observations show that this difference is easily differentiated with the NIRSpec G395H mode of {\it JWST}

We find that CO$_2$ is the cause of the difference at $\sim4.3~\mu$m. We performed an identical calculation (not shown) except that the opacity contribution of CO$_2$ was excluded, which results in a much smaller difference between the two spectra. For the model that increases O/H, the mole fraction of CO$_2$ is increased above that of the solar case (see \cref{figure:chemwarm}). This increases the opacity in the spectral regions that CO$_2$ absorbs and shifts the photosphere at these wavelengths to lower pressures. Since the temperature decreases with decreasing pressure this results in a cooler photospheric temperature and hence smaller atmospheric emission flux. On the other hand, for the calculation that decreases C/H the mole fraction of CO$_2$ is less than for the solar case, decreasing the opacity in the region that CO$_2$ absorbs and shifting the photosphere to higher pressures and temperatures. 

For the rest of the spectral range, which is largely dominated by H$_2$O absorption, there is little difference with the solar case despite the large changes in the H$_2$O abundance shown in \cref{figure:chemwarm}. This is due to changes in the $P$-$T$ profile mostly compensating for the changes in the H$_2$O abundance. To demonstrate this in \cref{figure:emisswarm2} we show the emission spectra of the hot atmosphere, as in \cref{figure:emisswarm}, except that we use the $P$-$T$ profile calculated using solar element abundances to calculate the chemical equilibrium abundances and emission spectrum. In other words, the temperature structure of the atmosphere is not consistent with the chemical composition. This is a common approach in the literature \citep[e.g.][]{MadMJ11,Kopparapu2012,Moses2013,Venot2015,TsaLG2017} where the chemical composition as a function of the C/O ratio is investigated with an assumed, constant $P$-$T$ profile, typically derived from solar element abundances. We refer to this as the ``non-consistent'' approach.

Comparing \cref{figure:emisswarm2} with \cref{figure:emisswarm} it is clear that there are much larger differences between the spectra for the calculations that vary O/H or vary C/H when the chemical composition and $P$-$T$ profile are calculated non-consistently. There are also larger differences between the C/O = 0.1 spectrum and the spectrum obtained assuming solar element abundances. Changes in the chemical composition, as the C/O ratio is varied, leads to a change in the atmospheric opacity. This in turn changes the pressure level and, potentially, the temperature of the photosphere. If the composition is solved consistently with the temperature profile, the $P$-$T$ profile can adjust to maintain radiative-convective equilibrium in response to these changes. However, if the $P$-$T$ profile is treated as a fixed model input then this is not possible and the resulting model atmosphere may no longer satisfy radiative-convective equilibrium. We will return to discuss this further in \cref{section:consistent}.

For C/O = 0.1, H$_2$O is the prominent absorber across much of the spectral range. For both calculations that increase O/H and that decrease C/H, the H$_2$O abundance increases, shifting the photosphere to lower pressures and, therefore, lower temperatures. For the inconsistent calculation this leads to a reduction of the flux ratio across almost all of the spectral range. The increase in the H$_2$O abundance is larger when O/H is increased with a fixed C/H compared with when C/H is decreased with a fixed O/H resulting in a more significant decrease in the flux ratio for the former case. 

The emission spectra for the hot atmosphere with C/O = 2 shown in \cref{figure:emisswarm}, calculated consistently with the $P$-$T$ profile, shows a significant difference with the solar case since CH$_4$ and H$_2$O effectively swap roles as the dominant absorber (\cref{figure:chemwarm}). Though the simulated spectra for the models that decrease O/H and that increase C/H look very similar by eye, a simple $\chi^2$-test reveals that there is a significant statistical difference between the two cases, when considered over such a broad wavelength scale. When the emission spectrum is calculated inconsistently with the $P$-$T$ profile (\cref{figure:emisswarm2}) larger differences are apparent, since the $P$-$T$ profile can no longer adjust to maintain radiative-convective equilibrium. Compared with the solar case, CH$_4$ replaces H$_2$O as the most important absorber across much of the spectral range. The CH$_4$ abundance is lower for the model that decreases O/H, compared with the model that increases C/H, meaning that the photosphere generally lies at higher pressures and temperatures for the former, resulting in a larger flux ratio.

\begin{figure*}
  \center
  \includegraphics[width=\textwidth]{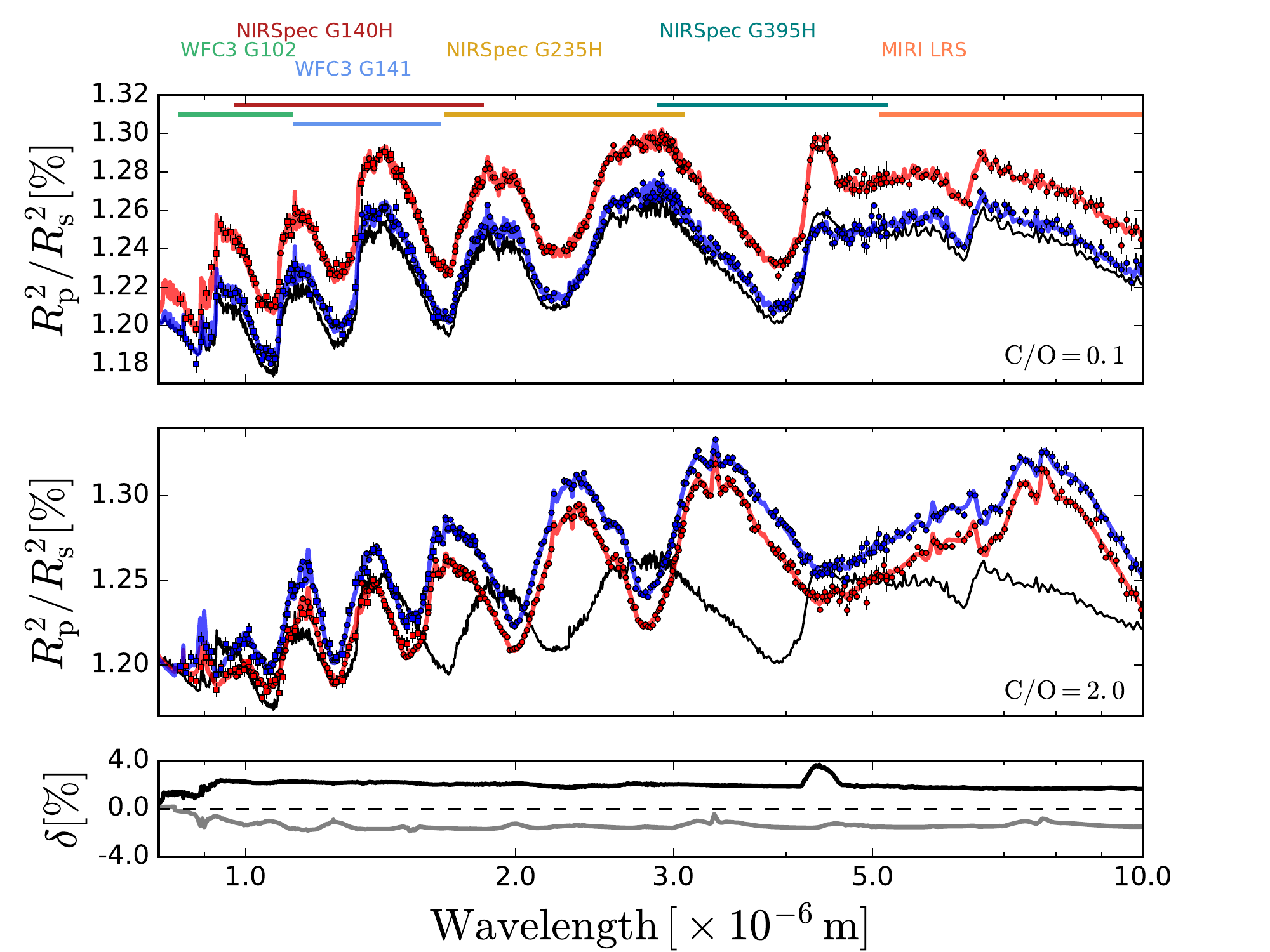}
   \caption{Transmission spectra for the hot atmosphere for C/O = 0.1 (top) and C/O = 2 (middle) for the model that varies O/H (red) and the model that varies C/H (blue). The transmission spectrum for the solar element abundances simulation (black) is also shown in both panels (top and middle). Also shown are PandExo simulated observations for the NIRSpec G140H, G235H and G395H and MIRI LRS modes ({\it JWST}, circles) as well as for WFC3 G102 and G141 ({\it HST}, squared), with the wavelength coverage of each instrument/mode indicated by the coloured horizontal lines in the top panel. The bottom panel shows the fractional difference between the models that vary O/H and the models that vary C/H for C/O = 0.1 (black) and C/O = 2 (grey). The black dashed line indicates a relative difference of $\delta = 0$. We note that we have shifted the spectra so that each is approximately aligned along the optical scattering slope.}
  \label{figure:transwarm}
\end{figure*}

\cref{figure:transwarm} shows the transmission spectra for the hot atmosphere. We note that in order to compare the spectra we have performed a vertical shift to approximately align each spectrum along the optical scattering slope. For C/O = 0.1 there is an overall shift in the transmission spectrum to larger transit depths for both calculations that increase O/H and that decrease C/H, compared to the solar case, though the shift is significantly larger for the former. Since the spectrum is dominated by H$_2$O absorption features, this is clearly mostly due to the larger mole fraction of H$_2$O that is obtained in the model that increases O/H, compared with the model that decreases C/H (\cref{figure:chemcool}). The relative difference between the models that increase O/H and that decrease C/H is $\sim2\%$ across a very large spectral range and the PandExo simulated observations, also shown in \cref{figure:transwarm}, clearly show that these differences are observationally significant with both current (i.e. {\it HST}) and future (i.e. {\it JWST}) instruments. The most notable difference in the spectra occurs at $\sim4.3~\mu$m. As previously identified in the emission spectrum (\cref{figure:emisswarm}) this is due to CO$_2$. The spectrum for the model that increases O/H shows a larger transit depth than the solar case, due to a larger CO$_2$ mole fraction, while the spectrum for the model that decreases C/H shows a smaller transit depth than the solar case, due to a smaller CO$_2$ mole fraction. The relative difference in the transit depth between the model approaches is $\sim4\%$ in this spectral region.

For C/O = 2 the spectra for both models that decrease O/H and that increase C/H are drastically different to the solar case, since CH$_4$ absorption features replace H$_2$O absorption features. Similar to the case at C/O = 0.1, there is a general shift in the transit depth, driven by the different abundances of CH$_4$ between the models that decrease O/H and that increase C/H (\cref{figure:chemwarm}), though the shapes of the spectra are very similar.

\subsubsection{Warm atmosphere}

\begin{figure*}
  \center
  \includegraphics[width=\textwidth]{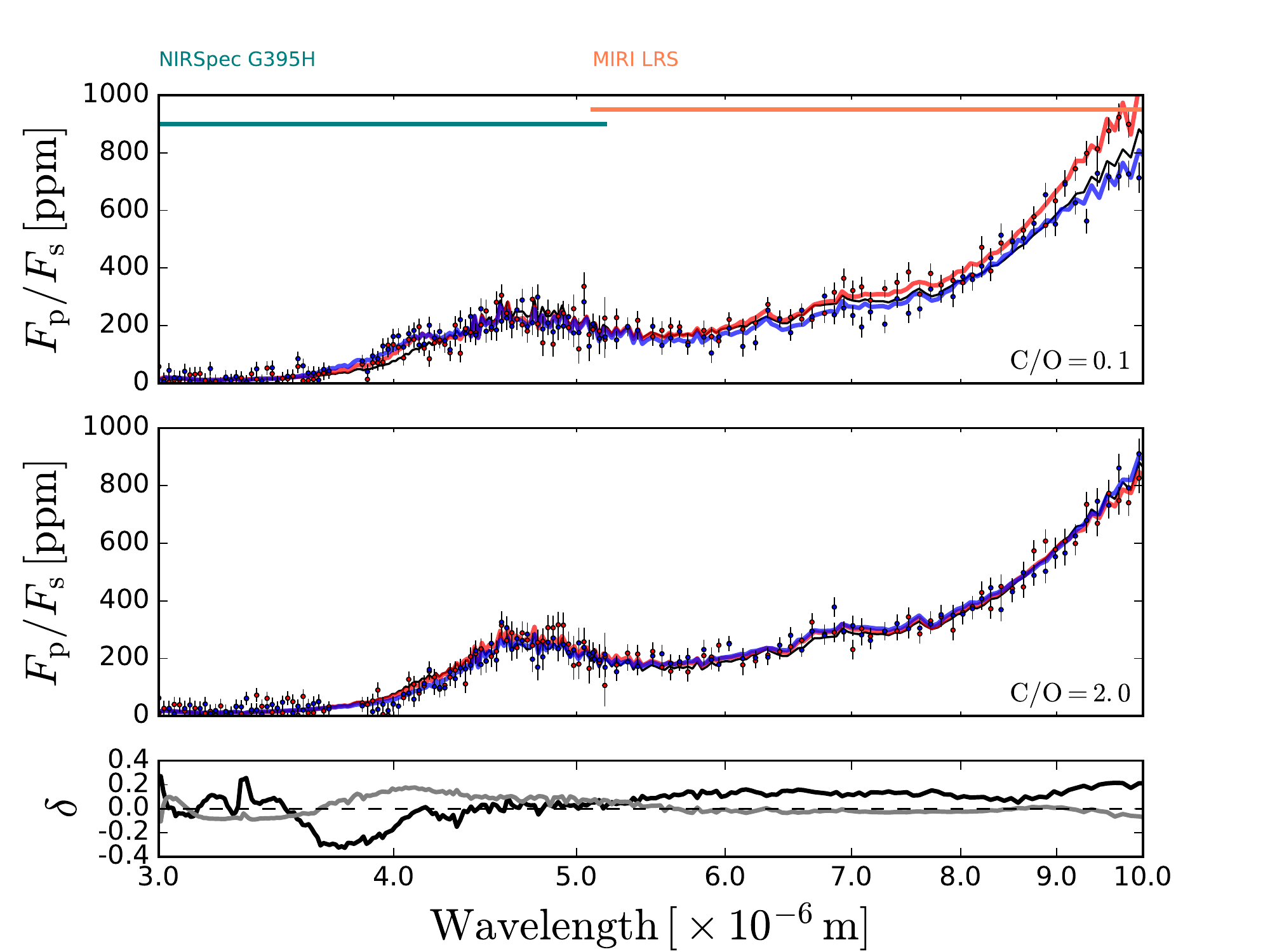}
   \caption{As \cref{figure:emisswarm} but for the warm atmosphere.}
  \label{figure:emisscool}
\end{figure*}

The emission spectra for the warm atmosphere, calculated using the consistent $P$-$T$ profile, are shown in \cref{figure:emisscool}. For both C/O = 0.1 and C/O = 2 there is generally only a small difference between the models that vary O/H and that vary C/H. Additionally, the spectra for the C/O = 0.1 and C/O = 2 cases show very little difference with the solar case. Though by eye the simulated observed spectra appear to show relatively insignificant differences, a $\chi^2$-test reveals that there is a significant statistical difference between the spectra from the calculations that vary O/H and the calculations that vary C/H. As for the hot atmosphere case this small difference in the spectra is largely due to changes in the $P$-$T$ profile partly compensating for changes in the chemical composition to maintain radiative-convective equilibrium.

\begin{figure*}
  \center
  \includegraphics[width=\textwidth]{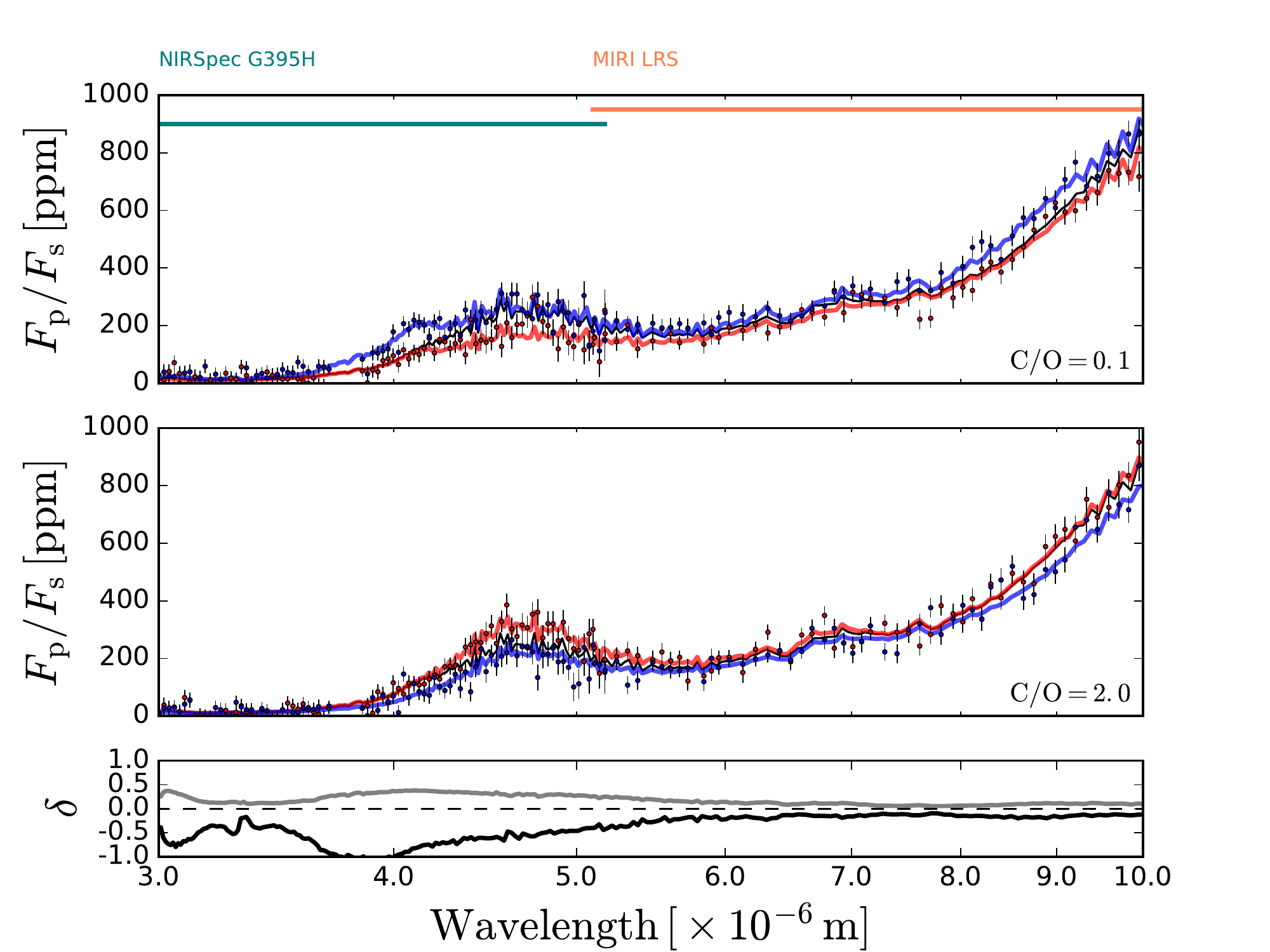}
   \caption{As \cref{figure:emisscool} but with the chemical composition and spectrum calculated using the $P$-$T$ profile calculated using solar element abundances: the non-consistent case.}
  \label{figure:emisscool2}
\end{figure*}

\cref{figure:emisscool2} shows the emission spectra for the warm atmosphere using the non-consistent approach; the $P$-$T$ profile is the same (the solar case) for all calculations. As previously found for the hot atmosphere, differences in the spectra between the solar and non-solar C/O ratio models, as well as between the models that vary O/H or C/H, are larger than for the consistent calculations (\cref{figure:emisscool}). The relative difference between the spectra for the models that vary O/H or C/H are as large as $\sim50\%$ for some wavelengths.

For C/O = 0.1 the model that increases O/H gives an emission spectrum with a smaller flux ratio for most wavelengths, particularly for 4 - 5 $\mu$m, compared with the model assuming solar element abundances. The major difference in the chemical composition are increases in the mole fractions of H$_2$O, CO and CO$_2$. These changes increase the opacity in the spectral regions where these molecules absorb and push the photosphere to a lower pressure and lower temperature, reducing the emitted flux. On the other hand, for the model that decreases C/H the main difference in the chemical composition is a decrease in the abundances of CH$_4$ CO and H$_2$O. This reduces the opacity in the spectral regions that these molecules absorb, shifting the photosphere to higher pressures and higher temperatures, increasing the emitted flux.

A similar situation is found for C/O = 2, though in the opposite sense. For the model that decreases O/H the abundances of H$_2$O, CO and CO$_2$ are decreased compared with the model assuming solar element abundances, leading to a larger emission flux for the wavelengths where these species absorb. On the other hand, for the model that increases C/H the abundances of CH$_4$, CO and CO$_2$ increase, decreasing the emission flux for the wavelengths where these species absorb.

\begin{figure*}
  \center
  \includegraphics[width=\textwidth]{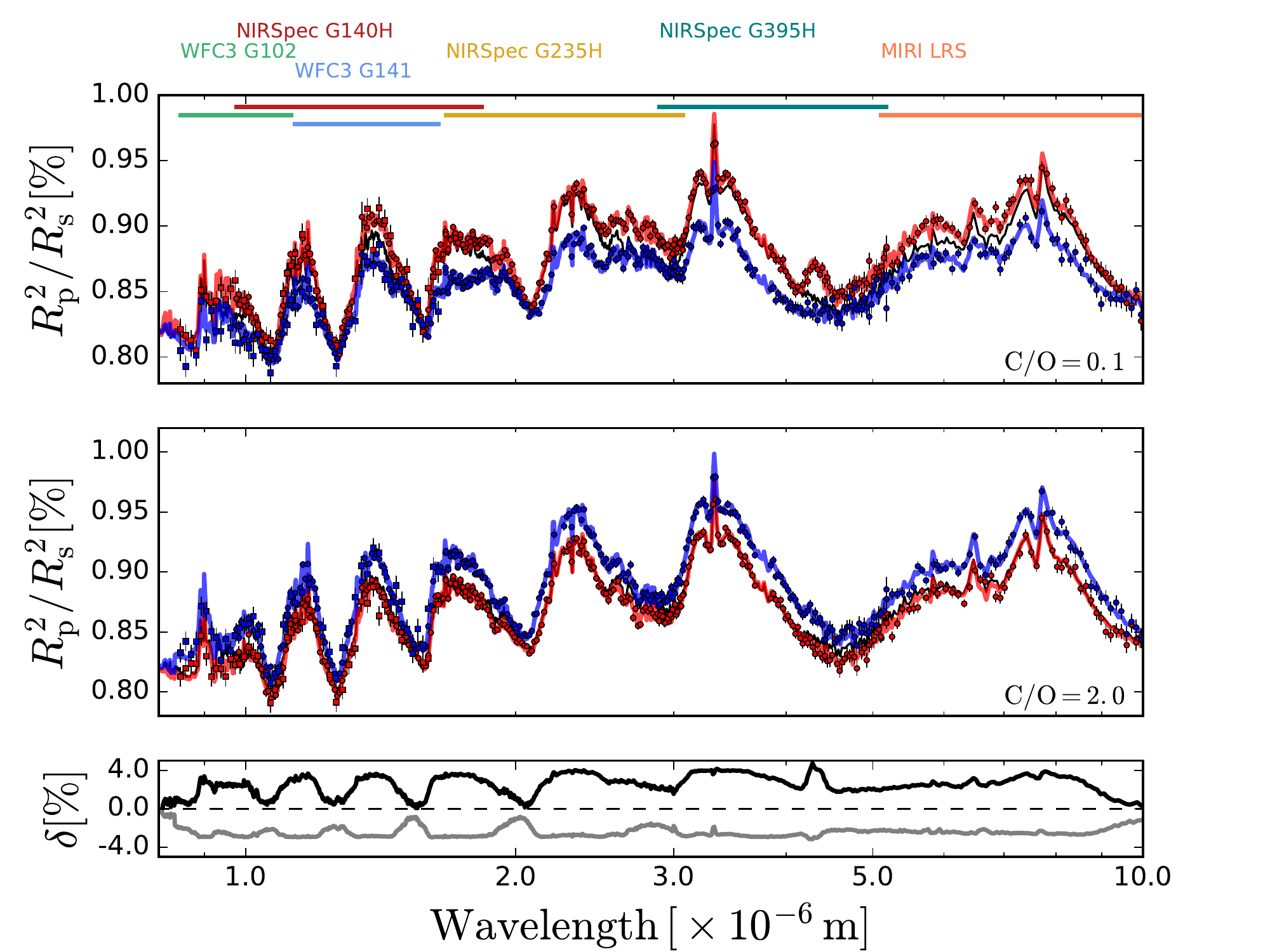}
   \caption{As \cref{figure:transwarm} but for the warm atmosphere.}
  \label{figure:transcool}
\end{figure*}

In \cref{figure:transcool} we show the transmission spectra for the warm atmosphere. There are significant quantitative and qualitative differences between the spectra for the models that vary O/H or C/H, for both C/O = 0.1 and C/O = 2. Relative differences in the transit depths between the models that vary O/H or C/H are as large as $\sim4\%$. The PandExo simulated observations demonstrate that differences in the spectra are clearly significant.

For C/O = 0.1, the transit depth around 1.4 $\mu$m is increased for the model that increases O/H, compared with the model assuming solar element abundances, while for the model that decreases C/H the transit depth decreases. H$_2$O and CH$_4$ are both important absorbers in this spectral region, which is accessible by both the WFC3 G141 ({\it HST}) and NIRSpec G140H ({\it JWST}) instruments. For the model that increases O/H the increased transit depth is due to the significant increase in the H$_2$O mole fraction (\cref{figure:chemcool}) compared with the solar case. On the other hand, the decrease in the transit depth for the model that decreases C/H is due to the significant decrease in the CH$_4$ mole fraction. The reduced CH$_4$ abundance in the spectrum for the model that reduces C/H is also notable in the smaller transit depths between 2-2.5 $\mu$m and 3-4 $\mu$m. Another notable feature lies at $\sim4.3~\mu$m where the spectrum for the model that increases O/H shows a significantly larger transit depth, compared to the spectra for the model that decreases C/H as well as the model that assumes solar element abundances, due to enhanced absorption of CO$_2$.

For C/O = 2, there is negligible difference between the spectra for the models that decrease O/H and that which assumes solar element abundances. Since the shape of the spectrum is largely dominated by CH$_4$ absorption features, this is because the abundance of CH$_4$ is virtually unchanged compared with the solar case (\cref{figure:chemcool}). However, for the model that increases C/H the spectrum is generally shifted to larger transit depths caused by the factor $\sim3$ increase in the mole fraction of CH$_4$, compared with the model that assumes solar element abundances.

\section{Discussion}
\label{section:discussion}
Our results have shown that the chemical composition, thermal structure and spectra predicted by models of hydrogen-dominated atmospheres for a given C/O ratio are dependent on the exact set of element abundances that are assumed. In this section we discuss some important implications of these results for previous and future studies. We then briefly discuss some works that investigate the processes and mechanisms that determine the element composition of exoplanet atmospheres.

\subsection{Implications for atmospheric forward modelling}
\label{section:implications}

\subsubsection{Chemical kinetics with 1D models}

The calculations performed in this paper assume local chemical equilibrium. However, it has been shown, using 1D chemical kinetics models, that vertical transport can be important in determining the abundances of radiatively-active (i.e. absorbing/scattering) chemical species, leading to changes in the simulated spectra \citep[e.g.][]{Moses2011,MilZF12,AguVS2014,Venot2015} and the temperature profile \citep{DruTB16}. Photochemistry has so far been shown to play a minor role in effecting the spectra.

Our results have implications for chemical kinetics studies that have investigated the effect of the C/O ratio \citep[e.g.][]{MadMJ11,Kopparapu2012,Moses2013,MosLV13,Venot2015,TsaLG2017}. We have shown that the chemical equilibrium composition depends on the method used to vary the C/O ratio, for a given $P$, $T$ and C/O ratio (\cref{section:box}). Therefore, for a given model $P$-$T$ profile, if vertical transport is included a different equilibrium composition at the pressure level of the quench point will likely exist which will lead to different quenched abundances at lower pressures. Additionally, we have shown that the $P$-$T$ profile for a given C/O ratio can be significantly different depending on whether the values of O/H and C/H that are assumed. Due to the temperature dependence of the chemical reactions, the pressure level of the quench point itself can be shifted, also leading to different quenched abundances at lower pressures.

\subsubsection{Three--dimensional (3D) modelling}

3D models of exoplanet atmospheres have been widely used to study the large--scale fluid flow of tidally--locked highly--irradiated planets \citep[e.g.][]{MenR09,Showman2009,DobA13,MayBA14,AmuMB16,MenGG16}. The vast majority of studies that include non-gray radiative transfer schemes assume local chemical equilibrium.

A number of studies have investigated the effect of non--solar element abundances on the circulation and thermal structure of irradiated exoplanet atmospheres, though these have focused on the metallicity \citep{Men12,KatSF14,ChaML15,ChaMM15,DruMB18}, as opposed to the C/O ratio. As far as we are aware, there has been no study to investigate the effect of different C/O ratios on the thermal structure and circulation using a 3D model. However, our results, using a 1D model, suggest different results may be obtained depending on exactly how the C/O ratio is varied.

Recently, \citet{MenTM18} investigated the effect of 3D advection on the chemical composition using a chemical relaxation scheme \citep{TsaKL18} for different C/O ratios, however they do not explicitly state how they vary the C/O ratio in their model. As in the case of 1D chemical models, a different chemical equilibrium composition for the same C/O ratio, due to variations in the assumed O/H and C/H, may lead to a different 3D chemical composition. If the chemistry is treated consistently with the radiative transfer, as in \citet[][]{DruMM18,DruMM18b}, this may also lead to differences in the circulation and thermal structure.

\subsubsection{Consistency between chemical composition and thermal structure}
\label{section:consistent}

In \cref{section:spectra} we presented emission spectra calculated using two different approaches. Firstly, we presented spectra where the radiative-convective equilibrium $P$-$T$ profile is solved for consistently with the chemical equilibrium composition for each C/O ratio case. Changing the C/O ratio leads to alterations in the $P$-$T$ profile due to changes in the atmospheric opacity. In our second approach, we calculated both the chemical composition and the emission spectrum using the $P$-$T$ profile obtained when assuming solar element abundances. If the abundances of absorbing species change significantly as the C/O ratio is varied, this can mean that the $P$-$T$ profile no longer satisfies radiative-convective equilibrium. This ``non-consistent'' approach is the typical approach in the literature \citep[e.g.][]{MadMJ11,Kopparapu2012,Moses2013,Venot2015,TsaLG2017}.

In general, we found much larger differences in the emission spectra between the solar and non-solar C/O cases, as well as between the models that vary O/H or C/H, for the non-consistent calculation compared with the calculation where the $P$-$T$ profile is consistent with the chemical composition (compare \cref{figure:emisswarm} with \cref{figure:emisswarm2} and \cref{figure:emisscool} with \cref{figure:emisscool2}). The reason behind this difference is analogous to the mechanism identified by \citet{DruTB16}, except that in this case changes to the chemical composition are due to changes in the element abundances, rather than due to non-equilibrium (vertical transport) effects.

Changing the element abundances (via the C/O ratio, for example) leads to a change in the equilibrium chemical composition and hence also the opacity of the atmosphere, if radiatively--active species are affected. If the $P$-$T$ profile is treated as a fixed model parameter, the atmosphere may no longer be in a state of radiative--convective equilibrium as the composition is varied. However, in the consistent models we solve for chemical and radiative--convective equilibrium consistently and the converged solution does, naturally, satisfy radiative--convective equilibrium. The larger difference between the spectra in the non--consistent calculations, compared with the consistent calculations, is therefore due to a violation of energy balance in the model. 

These results suggest that previous studies, using both chemical equilibrium and chemical kinetics, may have overpredicted the effect of changing the C/O ratio on the emission spectrum if the chemical composition and thermal structure were not solved for consistently.

\subsection{Implications for retrieval calculations and comparison of model spectra with observed spectra}
\label{section:retrievals}

Our results have clearly shown that differences in the predicted emission and transmission spectra can result from different approaches to varying the C/O ratio. This has obvious consequences for comparison of model spectra with observations and our results suggest care should be taken when the C/O ratio is considered as a variable model parameter.

Retrieval calculations combine simplified atmospheric models with statistical algorithms to derive the properties of an atmosphere from an observed spectrum \citep[e.g.][]{IrwTd08,MadS09,LinWZ13,WalTR15,EvaSK17,OreLG17,MarFS18,GanM18}. There is a significant variation in the literature in both the statistical methods and the setup of the atmosphere model component. 

In terms of the approach to the chemical composition, retrieval methods can be split into two main categories: ``free chemistry'' and chemical equilibrium. In the former, the mole fractions of the chemical species are included as independent free parameters, with quantities such as the C/O ratio and metallicity being derived from those retrieved abundances \citep[e.g.][]{LinWZ13,WalTR15,GanM18}. On the other hand, for the chemical equilibrium method, the C/O ratio and metallicity are typically directly included as retrieved parameters \citep[e.g.][]{KreLB15,LavMM17,WakSD18}. \citet{OreLG17} include C/H and O/H as independent parameters, rather than the more typical combination of metallicity and C/O ratio.

Retrieval calculations that assume the ``free chemistry'' approach are not affected by our findings since there is no need to make a decision on how to vary the C/O ratio in the model. The C/O ratio is simply determined from the best-fit mole fractions of the included chemical species \citep[e.g.][]{LinWZ13}. On the other hand, retrieval calculations that assume local chemical equilibrium can be dependent on whether O/H or C/H is varied. \citet{WakSD18}, who use the same forward model (ATMO) in their retrieval calculations as used in this work, clearly state that for their calculations assuming local chemical equilibrium C/O is varied by adjusting the value of O/H. 

We have shown that the predicted spectrum can be dependent on exactly how the C/O ratio is varied in the forward model. Therefore, it is reasonable to assume that these differences could effect retrieved parameters, for retrieval calculations that assume local chemical equilibrium. The method of \citet{OreLG17}, who include individual parameters for O/H and C/H in their retrieval calculation, may be a better approach as it would avoid the need to make a choice on how to vary the C/O ratio in the model.

\subsection{C/O of planetary atmospheres}
\label{section:co}

The main finding of this paper is that different atmospheric compositions, thermal structures and spectra are obtained for the same C/O ratios and metallicities depending on whether we vary C/H for a given O/H or vary O/H for a given C/H. Here, we briefly discuss some previous results that may be relevant to answering the question of which, if either, of these approaches is more appropriate.

The measured C/O ratio of stars, including planet-hosting stars, is known to vary and the C/O ratio is generally observed to increase with increasing stellar metallicity \citep[e.g.][]{DelIG10,PetM11,Nis13,TesCS14,BreF16,BreFM17,NisG18,JofHS18}; see \citet{NisG18} for a recent review of stellar element abundances. Specifically, the C/O ratio is shown to increase with increasing O/H for [O/H] $>-0.4$ \citep[e.g.][]{AkeCN04,NisCC14}, as carbon becomes proportionately more abundant than oxygen. Assuming that planets form from the same material as the star, the varied chemical composition observed in the stellar population is likely to be reflected in the composition of the planetary population.

Processes in the protoplanetary disk and the mechanism of planet formation can also lead to different element compositions of the planet. \citet{ObeMB11} used a static disk model to demonstrate that gas in regions of the protoplanetary disk should be enhanced in carbon, due to condensation of H$_2$O, CO and CO$_2$; i.e. the gas is predominantly carbon-rich, with respect to the stellar composition, with oxygen-rich solids. The planetary C/O ratio would then depend on the composition of the disk in the region of planet formation, as well as subsequent accretion of either (oxygen-rich) icy solids or carbon-rich material. \citet{PisOB15} expanded on this work by included radial drift and viscous gas accretion and came to similar conclusions.  \citet{MorvM16} and \citet{EspFM17} found that accretion of solid material is the dominant factor determining the planetary C/O ratio and predicted that almost all planets should have substellar C/O ratios, assuming that accreted solids are oxygen-rich. \citet{MadBJ17} combined a planet growth and chemistry model and found that planets with superstellar C/O ratios (with stellar C/H and slightly substellar O/H) are possible for hot Jupiters that migrate through the disk and do not undergo core erosion. If core erosion occurs, subsequent mixing into the planetary envelope can result in superstellar O/H and C/H and a substellar C/O ratio. 

\citet{HelWR14}, using a time-dependent disk and chemistry evolution code, suggested that cosmic ray ionisation can increase the gas phase C/O ratio of the disk over time, within the CO ice-line, as species such as O$_2$ and CO are converted into H$_2$O which can then condense, reducing the O/H of the gas. Recently, \citet{EisWv18} also used a disk and chemical evolution model to show that the element abundances in both gas and ice are strongly time-dependent, largely driven by ionisation. They find that the gas-phase C/O ratio decreases significantly over time, largely due to significant production of gas-phase O$_2$ via grain surface reactions \citep{WalNv15}.

Aside from the bulk element chemical composition of the planet, a result of the complex formation and evolution processes discussed above, chemistry within the planetary atmosphere itself can alter the gas-phase C/O. Namely, condensation of silicon and magnesium condensates, that contain many oxygen atoms, can remove significant amounts of oxygen from the gas-phase \citep[e.g.][]{Burrows1999,Lod04,WoiHH18}, reducing O/H in the upper atmosphere. 

From the above discussion, it is clear that understanding the element composition of the gaseous and solid material that eventually form a planet is a very active field, with many open questions. The relative importance of the accretion of gas or solids in determining the chemical composition is also currently unclear. We conclude that, at present, the bulk element composition of exoplanets (O/H, C/H, N/H etc.) is highly unconstrained.

\section{Conclusions}
\label{section:conclusions}

We have demonstrated that different approaches to varying the C/O ratio can lead to significant differences in the calculated atmospheric properties, including the chemical equilibrium composition, the thermal profile and the emission and transmission spectra. The typical approach in the literature has so far been to either vary the carbon abundance while keeping the oxygen abundance fixed or to vary the oxygen abundance while keeping the carbon abundance fixed. Importantly, the processes and mechanisms that determine the element composition of exoplanets and their atmospheres are highly uncertain and are a very active area of research, as discussed in \cref{section:co}.

To summarise our results, we have shown that: 
\begin{enumerate}
  \item{the chemical equilibrium composition for the same C/O ratio can be significantly different depending on the assumed values of C/H and O/H, as previously shown by \citet{MosLV13}, \citet{HenL16} and \citet{HenT16}}
  \item{differences in chemical composition for the same C/O ratio but different C/H and O/H can result in significantly different mean molecular masses for the gas mixture}
  \item{using a 1D atmosphere model, differences in the equilibrium composition can lead to significant differences in the radiative--convective equilibrium $P$-$T$ profile}
  \item{differences in both the chemical composition and $P$-$T$ profile can lead to significant differences in the transmission and emission spectrum, which are detectable with current and future instruments}
  \item{keeping the $P$-$T$ profile fixed (i.e. not consistent with the chemistry) while investigating different C/O ratios can significantly overestimate the importance of the C/O ratio on the emission spectrum compared with a model where the chemistry and the temperature are solved consistently}
\end{enumerate}

In this work we have focussed on the C/O ratio, but our conclusions are also likely to be applicable to other element ratios (e.g. C/N, O/N). However, the C/O ratio appears to be particularly important since carbon and oxygen are among the most abundant elements in the universe and form relatively abundant, highly absorbing gas-phase species (e.g. H$_2$O, CH$_4$, CO, CO$_2$) in planetary atmospheres. 

We have discussed potential implications of our results, including: the quenching point of chemical species in 1D chemical kinetics models, the thermal structure and circulation predicted by 3D atmospheric models and retrieved parameters from retrieval codes that assume local chemical equilibrium. Our results show that it is important to consider the full set of element abundances (e.g. C/H, O/H, N/H) rather than individual ratios of element abundances (e.g. C/O) in isolation when modelling or interpreting the observations of irradiated, hydrogen-dominated atmospheres.

\section*{Acknowledgements}
The authors thank the anonymous referee whose comments greatly improved the quality of this manuscript. The authors also thank Isabelle Baraffe for comments on an earlier draft of this manuscript. BD, DKS and TME acknowledge support from the European Research Council under the
European Community's Seventh Framework Programme (FP7/2007-2013 Grant Agreement No. 336792-CREATES). BD, EH and NJM acknowledge support from a Science and Technology Facilities Council Consolidated Grant (ST/R000395/1). ALC is funded by an STFC studentship. JG and NJM are partially funded by a Leverhulme Trust Research Project Grant.
%




\bibliographystyle{mnras}

\begin{thebibliography}{}
\makeatletter
\relax
\def\mn@urlcharsother{\let\do\@makeother \do\$\do\&\do\#\do\^\do\_\do\%\do\~}
\def\mn@doi{\begingroup\mn@urlcharsother \@ifnextchar [ {\mn@doi@}
  {\mn@doi@[]}}
\def\mn@doi@[#1]#2{\def\@tempa{#1}\ifx\@tempa\@empty \href
  {http://dx.doi.org/#2} {doi:#2}\else \href {http://dx.doi.org/#2} {#1}\fi
  \endgroup}
\def\mn@eprint#1#2{\mn@eprint@#1:#2::\@nil}
\def\mn@eprint@arXiv#1{\href {http://arxiv.org/abs/#1} {{\tt arXiv:#1}}}
\def\mn@eprint@dblp#1{\href {http://dblp.uni-trier.de/rec/bibtex/#1.xml}
  {dblp:#1}}
\def\mn@eprint@#1:#2:#3:#4\@nil{\def\@tempa {#1}\def\@tempb {#2}\def\@tempc
  {#3}\ifx \@tempc \@empty \let \@tempc \@tempb \let \@tempb \@tempa \fi \ifx
  \@tempb \@empty \def\@tempb {arXiv}\fi \@ifundefined
  {mn@eprint@\@tempb}{\@tempb:\@tempc}{\expandafter \expandafter \csname
  mn@eprint@\@tempb\endcsname \expandafter{\@tempc}}}

\bibitem[\protect\citeauthoryear{{Ag{\'u}ndez}, {Venot}, {Selsis}  \&
  {Iro}}{{Ag{\'u}ndez} et~al.}{2014}]{AguVS2014}
{Ag{\'u}ndez} M.,  {Venot} O.,  {Selsis} F.,   {Iro} N.,  2014, \mn@doi [\apj]
  {10.1088/0004-637X/781/2/68}, \href
  {http://adsabs.harvard.edu/abs/2014ApJ...781...68A} {781, 68}

\bibitem[\protect\citeauthoryear{{Akerman}, {Carigi}, {Nissen}, {Pettini}  \&
  {Asplund}}{{Akerman} et~al.}{2004}]{AkeCN04}
{Akerman} C.~J.,  {Carigi} L.,  {Nissen} P.~E.,  {Pettini} M.,   {Asplund} M.,
  2004, \mn@doi [\aap] {10.1051/0004-6361:20034188}, \href
  {http://adsabs.harvard.edu/abs/2004A%26A...414..931A} {414, 931}

\bibitem[\protect\citeauthoryear{{Allard}, {Homeier}, {Freytag}  \&
  {Sharp}}{{Allard} et~al.}{2012}]{Allard2012}
{Allard} F.,  {Homeier} D.,  {Freytag} B.,   {Sharp} C.~M.,  2012, in
  {Reyl{\'e}} C.,  {Charbonnel} C.,   {Schultheis} M.,  eds,  EAS Publications
  Series Vol. 57, EAS Publications Series. pp 3--43 (\mn@eprint {arXiv}
  {1206.1021}), \mn@doi{10.1051/eas/1257001}

\bibitem[\protect\citeauthoryear{{Allende Prieto}}{{Allende
  Prieto}}{2016}]{Pri16}
{Allende Prieto} C.,  2016, \mn@doi [Living Reviews in Solar Physics]
  {10.1007/s41116-016-0001-6}, \href
  {http://adsabs.harvard.edu/abs/2016LRSP...13....1A} {13, 1}

\bibitem[\protect\citeauthoryear{{Amundsen} et~al.,}{{Amundsen}
  et~al.}{2016}]{AmuMB16}
{Amundsen} D.~S.,  et~al., 2016, \mn@doi [\aap] {10.1051/0004-6361/201629183},
  \href {http://adsabs.harvard.edu/abs/2016A%26A...595A..36A} {595, A36}

\bibitem[\protect\citeauthoryear{{Amundsen}, {Tremblin}, {Manners}, {Baraffe}
  \& {Mayne}}{{Amundsen} et~al.}{2017}]{AmuTM17}
{Amundsen} D.~S.,  {Tremblin} P.,  {Manners} J.,  {Baraffe} I.,   {Mayne}
  N.~J.,  2017, \mn@doi [\aap] {10.1051/0004-6361/201629322}, \href
  {http://adsabs.harvard.edu/abs/2017A%26A...598A..97A} {598, A97}

\bibitem[\protect\citeauthoryear{{Asplund}, {Grevesse}, {Sauval}  \&
  {Scott}}{{Asplund} et~al.}{2009}]{Asplund2009}
{Asplund} M.,  {Grevesse} N.,  {Sauval} A.~J.,   {Scott} P.,  2009, \mn@doi
  [\araa] {10.1146/annurev.astro.46.060407.145222}, \href
  {http://adsabs.harvard.edu/abs/2009ARA%26A..47..481A} {47, 481}

\bibitem[\protect\citeauthoryear{{Batalha} et~al.,}{{Batalha}
  et~al.}{2017}]{BatMP17}
{Batalha} N.~E.,  et~al., 2017, \mn@doi [\pasp] {10.1088/1538-3873/aa65b0},
  \href {http://adsabs.harvard.edu/abs/2017PASP..129f4501B} {129, 064501}

\bibitem[\protect\citeauthoryear{{Brewer} \& {Fischer}}{{Brewer} \&
  {Fischer}}{2016}]{BreF16}
{Brewer} J.~M.,  {Fischer} D.~A.,  2016, \mn@doi [\apj]
  {10.3847/0004-637X/831/1/20}, \href
  {http://adsabs.harvard.edu/abs/2016ApJ...831...20B} {831, 20}

\bibitem[\protect\citeauthoryear{{Brewer}, {Fischer}  \&
  {Madhusudhan}}{{Brewer} et~al.}{2017}]{BreFM17}
{Brewer} J.~M.,  {Fischer} D.~A.,   {Madhusudhan} N.,  2017, \mn@doi [\aj]
  {10.3847/1538-3881/153/2/83}, \href
  {http://adsabs.harvard.edu/abs/2017AJ....153...83B} {153, 83}

\bibitem[\protect\citeauthoryear{{Burrows} \& {Sharp}}{{Burrows} \&
  {Sharp}}{1999}]{Burrows1999}
{Burrows} A.,  {Sharp} C.~M.,  1999, \mn@doi [\apj] {10.1086/306811}, \href
  {http://adsabs.harvard.edu/abs/1999ApJ...512..843B} {512, 843}

\bibitem[\protect\citeauthoryear{{Caffau}, {Ludwig}, {Steffen}, {Freytag}  \&
  {Bonifacio}}{{Caffau} et~al.}{2011}]{Caffau2011}
{Caffau} E.,  {Ludwig} H.-G.,  {Steffen} M.,  {Freytag} B.,   {Bonifacio} P.,
  2011, \mn@doi [\solphys] {10.1007/s11207-010-9541-4}, \href
  {http://adsabs.harvard.edu/abs/2011SoPh..268..255C} {268, 255}

\bibitem[\protect\citeauthoryear{{Charnay}, {Meadows}  \& {Leconte}}{{Charnay}
  et~al.}{2015a}]{ChaML15}
{Charnay} B.,  {Meadows} V.,   {Leconte} J.,  2015a, \mn@doi [\apj]
  {10.1088/0004-637X/813/1/15}, \href
  {http://adsabs.harvard.edu/abs/2015ApJ...813...15C} {813, 15}

\bibitem[\protect\citeauthoryear{{Charnay}, {Meadows}, {Misra}, {Leconte}  \&
  {Arney}}{{Charnay} et~al.}{2015b}]{ChaMM15}
{Charnay} B.,  {Meadows} V.,  {Misra} A.,  {Leconte} J.,   {Arney} G.,  2015b,
  \mn@doi [\apjl] {10.1088/2041-8205/813/1/L1}, \href
  {http://adsabs.harvard.edu/abs/2015ApJ...813L...1C} {813, L1}

\bibitem[\protect\citeauthoryear{{Delgado Mena}, {Israelian}, {Gonz{\'a}lez
  Hern{\'a}ndez}, {Bond}, {Santos}, {Udry}  \& {Mayor}}{{Delgado Mena}
  et~al.}{2010}]{DelIG10}
{Delgado Mena} E.,  {Israelian} G.,  {Gonz{\'a}lez Hern{\'a}ndez} J.~I.,
  {Bond} J.~C.,  {Santos} N.~C.,  {Udry} S.,   {Mayor} M.,  2010, \mn@doi
  [\apj] {10.1088/0004-637X/725/2/2349}, \href
  {http://adsabs.harvard.edu/abs/2010ApJ...725.2349D} {725, 2349}

\bibitem[\protect\citeauthoryear{{Deming} et~al.,}{{Deming}
  et~al.}{2013}]{Deming2013}
{Deming} D.,  et~al., 2013, \mn@doi [\apj] {10.1088/0004-637X/774/2/95}, \href
  {http://adsabs.harvard.edu/abs/2013ApJ...774...95D} {774, 95}

\bibitem[\protect\citeauthoryear{{Dobbs-Dixon} \& {Agol}}{{Dobbs-Dixon} \&
  {Agol}}{2013}]{DobA13}
{Dobbs-Dixon} I.,  {Agol} E.,  2013, \mn@doi [\mnras] {10.1093/mnras/stt1509},
  \href {http://adsabs.harvard.edu/abs/2013MNRAS.435.3159D} {435, 3159}

\bibitem[\protect\citeauthoryear{{Drummond}, {Tremblin}, {Baraffe}, {Amundsen},
  {Mayne}, {Venot}  \& {Goyal}}{{Drummond} et~al.}{2016}]{DruTB16}
{Drummond} B.,  {Tremblin} P.,  {Baraffe} I.,  {Amundsen} D.~S.,  {Mayne}
  N.~J.,  {Venot} O.,   {Goyal} J.,  2016, \mn@doi [\aap]
  {10.1051/0004-6361/201628799}, \href
  {http://adsabs.harvard.edu/abs/2016A%26A...594A..69D} {594, A69}

\bibitem[\protect\citeauthoryear{{Drummond}, {Mayne}, {Baraffe}, {Tremblin},
  {Manners}, {Amundsen}, {Goyal}  \& {Acreman}}{{Drummond}
  et~al.}{2018a}]{DruMB18}
{Drummond} B.,  {Mayne} N.~J.,  {Baraffe} I.,  {Tremblin} P.,  {Manners} J.,
  {Amundsen} D.~S.,  {Goyal} J.,   {Acreman} D.,  2018a, \mn@doi [\aap]
  {10.1051/0004-6361/201732010}, \href
  {http://adsabs.harvard.edu/abs/2018A%26A...612A.105D} {612, A105}

\bibitem[\protect\citeauthoryear{{Drummond} et~al.,}{{Drummond}
  et~al.}{2018b}]{DruMM18}
{Drummond} B.,  et~al., 2018b, \mn@doi [\apjl] {10.3847/2041-8213/aab209},
  \href {http://adsabs.harvard.edu/abs/2018ApJ...855L..31D} {855, L31}

\bibitem[\protect\citeauthoryear{{Drummond}, {Mayne}, {Manners}, {Baraffe},
  {Goyal}, {Tremblin}, {Sing}  \& {Kohary}}{{Drummond}
  et~al.}{2018c}]{DruMM18b}
{Drummond} B.,  {Mayne} N.~J.,  {Manners} J.,  {Baraffe} I.,  {Goyal} J.,
  {Tremblin} P.,  {Sing} D.~K.,   {Kohary} K.,  2018c, \mn@doi [\apj]
  {10.3847/1538-4357/aaeb28}, \href
  {http://adsabs.harvard.edu/abs/2018ApJ...869...28D} {869, 28}

\bibitem[\protect\citeauthoryear{{Eistrup}, {Walsh}  \& {van
  Dishoeck}}{{Eistrup} et~al.}{2018}]{EisWv18}
{Eistrup} C.,  {Walsh} C.,   {van Dishoeck} E.~F.,  2018, \mn@doi [\aap]
  {10.1051/0004-6361/201731302}, \href
  {http://adsabs.harvard.edu/abs/2018A%26A...613A..14E} {613, A14}

\bibitem[\protect\citeauthoryear{{Espinoza}, {Fortney}, {Miguel}, {Thorngren}
  \& {Murray-Clay}}{{Espinoza} et~al.}{2017}]{EspFM17}
{Espinoza} N.,  {Fortney} J.~J.,  {Miguel} Y.,  {Thorngren} D.,   {Murray-Clay}
  R.,  2017, \mn@doi [\apjl] {10.3847/2041-8213/aa65ca}, \href
  {http://adsabs.harvard.edu/abs/2017ApJ...838L...9E} {838, L9}

\bibitem[\protect\citeauthoryear{{Evans} et~al.,}{{Evans}
  et~al.}{2017}]{EvaSK17}
{Evans} T.~M.,  et~al., 2017, \mn@doi [\nat] {10.1038/nature23266}, \href
  {http://adsabs.harvard.edu/abs/2017Natur.548...58E} {548, 58}

\bibitem[\protect\citeauthoryear{{Fegley} \& {Lodders}}{{Fegley} \&
  {Lodders}}{1996}]{FegL96}
{Fegley} Jr. B.,  {Lodders} K.,  1996, \mn@doi [\apjl] {10.1086/310356}, \href
  {http://adsabs.harvard.edu/abs/1996ApJ...472L..37F} {472, L37}

\bibitem[\protect\citeauthoryear{{Fortney}, {Marley}, {Lodders}, {Saumon}  \&
  {Freedman}}{{Fortney} et~al.}{2005}]{ForML05}
{Fortney} J.~J.,  {Marley} M.~S.,  {Lodders} K.,  {Saumon} D.,   {Freedman} R.,
   2005, \mn@doi [\apjl] {10.1086/431952}, \href
  {http://adsabs.harvard.edu/abs/2005ApJ...627L..69F} {627, L69}

\bibitem[\protect\citeauthoryear{{Gaidos}}{{Gaidos}}{2000}]{Gai00}
{Gaidos} E.~J.,  2000, \mn@doi [\icarus] {10.1006/icar.2000.6407}, \href
  {http://adsabs.harvard.edu/abs/2000Icar..145..637G} {145, 637}

\bibitem[\protect\citeauthoryear{{Gandhi} \& {Madhusudhan}}{{Gandhi} \&
  {Madhusudhan}}{2017}]{GanM17}
{Gandhi} S.,  {Madhusudhan} N.,  2017, \mn@doi [\mnras]
  {10.1093/mnras/stx1601}, \href
  {http://adsabs.harvard.edu/abs/2017MNRAS.472.2334G} {472, 2334}

\bibitem[\protect\citeauthoryear{{Gandhi} \& {Madhusudhan}}{{Gandhi} \&
  {Madhusudhan}}{2018}]{GanM18}
{Gandhi} S.,  {Madhusudhan} N.,  2018, \mn@doi [\mnras]
  {10.1093/mnras/stx2748}, \href
  {http://adsabs.harvard.edu/abs/2018MNRAS.474..271G} {474, 271}

\bibitem[\protect\citeauthoryear{{Gordon} \& {McBride}}{{Gordon} \&
  {McBride}}{1994}]{Gordon1994}
{Gordon} S.,  {McBride} B.~J.,  1994, NASA Reference Publication, 1311

\bibitem[\protect\citeauthoryear{{Goyal} et~al.,}{{Goyal}
  et~al.}{2018}]{GoyMS18}
{Goyal} J.~M.,  et~al., 2018, \mn@doi [\mnras] {10.1093/mnras/stx3015}, \href
  {http://adsabs.harvard.edu/abs/2018MNRAS.474.5158G} {474, 5158}

\bibitem[\protect\citeauthoryear{{Grevesse} \& {Sauval}}{{Grevesse} \&
  {Sauval}}{1998}]{GreS98}
{Grevesse} N.,  {Sauval} A.~J.,  1998, \mn@doi [\ssr]
  {10.1023/A:1005161325181}, \href
  {http://adsabs.harvard.edu/abs/1998SSRv...85..161G} {85, 161}

\bibitem[\protect\citeauthoryear{{Helling} \& {Lucas}}{{Helling} \&
  {Lucas}}{2009}]{HelL09}
{Helling} C.,  {Lucas} W.,  2009, \mn@doi [\mnras]
  {10.1111/j.1365-2966.2009.15164.x}, \href
  {http://adsabs.harvard.edu/abs/2009MNRAS.398..985H} {398, 985}

\bibitem[\protect\citeauthoryear{{Helling}, {Woitke}, {Rimmer}, {Kamp}, {Thi}
  \& {Meijerink}}{{Helling} et~al.}{2014}]{HelWR14}
{Helling} C.,  {Woitke} P.,  {Rimmer} P.~B.,  {Kamp} I.,  {Thi} W.-F.,
  {Meijerink} R.,  2014, \mn@doi [Life] {10.3390/life4020142}, \href
  {http://adsabs.harvard.edu/abs/2014Life....4..142H} {4}

\bibitem[\protect\citeauthoryear{{Heng} \& {Lyons}}{{Heng} \&
  {Lyons}}{2016}]{HenL16}
{Heng} K.,  {Lyons} J.~R.,  2016, \mn@doi [\apj] {10.3847/0004-637X/817/2/149},
  \href {http://adsabs.harvard.edu/abs/2016ApJ...817..149H} {817, 149}

\bibitem[\protect\citeauthoryear{{Heng} \& {Tsai}}{{Heng} \&
  {Tsai}}{2016}]{HenT16}
{Heng} K.,  {Tsai} S.-M.,  2016, \mn@doi [\apj] {10.3847/0004-637X/829/2/104},
  \href {http://adsabs.harvard.edu/abs/2016ApJ...829..104H} {829, 104}

\bibitem[\protect\citeauthoryear{{Irwin} et~al.,}{{Irwin}
  et~al.}{2008}]{IrwTd08}
{Irwin} P.~G.~J.,  et~al., 2008, \mn@doi [\jqsrt]
  {10.1016/j.jqsrt.2007.11.006}, \href
  {http://adsabs.harvard.edu/abs/2008JQSRT.109.1136I} {109, 1136}

\bibitem[\protect\citeauthoryear{{Jofr{\'e}}, {Heiter}  \&
  {Soubiran}}{{Jofr{\'e}} et~al.}{2018}]{JofHS18}
{Jofr{\'e}} P.,  {Heiter} U.,   {Soubiran} C.,  2018, preprint, \href
  {http://adsabs.harvard.edu/abs/2018arXiv181108041J} {} (\mn@eprint {arXiv}
  {1811.08041})

\bibitem[\protect\citeauthoryear{{Kataria}, {Showman}, {Fortney}, {Marley}  \&
  {Freedman}}{{Kataria} et~al.}{2014}]{KatSF14}
{Kataria} T.,  {Showman} A.~P.,  {Fortney} J.~J.,  {Marley} M.~S.,   {Freedman}
  R.~S.,  2014, \mn@doi [\apj] {10.1088/0004-637X/785/2/92}, \href
  {http://adsabs.harvard.edu/abs/2014ApJ...785...92K} {785, 92}

\bibitem[\protect\citeauthoryear{{Kitzmann} et~al.,}{{Kitzmann}
  et~al.}{2018}]{KitHR18}
{Kitzmann} D.,  et~al., 2018, \mn@doi [\apj] {10.3847/1538-4357/aace5a}, \href
  {http://adsabs.harvard.edu/abs/2018ApJ...863..183K} {863, 183}

\bibitem[\protect\citeauthoryear{{Knutson}, {Benneke}, {Deming}  \&
  {Homeier}}{{Knutson} et~al.}{2014}]{KnuBD14}
{Knutson} H.~A.,  {Benneke} B.,  {Deming} D.,   {Homeier} D.,  2014, \mn@doi
  [\nat] {10.1038/nature12887}, \href
  {http://adsabs.harvard.edu/abs/2014Natur.505...66K} {505, 66}

\bibitem[\protect\citeauthoryear{{Konopacky}, {Barman}, {Macintosh}  \&
  {Marois}}{{Konopacky} et~al.}{2013}]{KonBM13}
{Konopacky} Q.~M.,  {Barman} T.~S.,  {Macintosh} B.~A.,   {Marois} C.,  2013,
  \mn@doi [Science] {10.1126/science.1232003}, \href
  {http://adsabs.harvard.edu/abs/2013Sci...339.1398K} {339, 1398}

\bibitem[\protect\citeauthoryear{{Kopparapu}, {Kasting}  \&
  {Zahnle}}{{Kopparapu} et~al.}{2012}]{Kopparapu2012}
{Kopparapu} R.~k.,  {Kasting} J.~F.,   {Zahnle} K.~J.,  2012, \mn@doi [\apj]
  {10.1088/0004-637X/745/1/77}, \href
  {http://adsabs.harvard.edu/abs/2012ApJ...745...77K} {745, 77}

\bibitem[\protect\citeauthoryear{{Kreidberg} et~al.,}{{Kreidberg}
  et~al.}{2015}]{KreLB15}
{Kreidberg} L.,  et~al., 2015, \mn@doi [\apj] {10.1088/0004-637X/814/1/66},
  \href {http://adsabs.harvard.edu/abs/2015ApJ...814...66K} {814, 66}

\bibitem[\protect\citeauthoryear{{Kuchner} \& {Seager}}{{Kuchner} \&
  {Seager}}{2005}]{KucS05}
{Kuchner} M.~J.,  {Seager} S.,  2005, arXiv Astrophysics e-prints, \href
  {http://adsabs.harvard.edu/abs/2005astro.ph..4214K} {}

\bibitem[\protect\citeauthoryear{{Lacis} \& {Oinas}}{{Lacis} \&
  {Oinas}}{1991}]{Lacis1991}
{Lacis} A.~A.,  {Oinas} V.,  1991, \mn@doi [\jgr] {10.1029/90JD01945}, \href
  {http://adsabs.harvard.edu/abs/1991JGR....96.9027L} {96, 9027}

\bibitem[\protect\citeauthoryear{{Lavie} et~al.,}{{Lavie}
  et~al.}{2017}]{LavMM17}
{Lavie} B.,  et~al., 2017, \mn@doi [\aj] {10.3847/1538-3881/aa7ed8}, \href
  {http://adsabs.harvard.edu/abs/2017AJ....154...91L} {154, 91}

\bibitem[\protect\citeauthoryear{{Lee}, {Heng}  \& {Irwin}}{{Lee}
  et~al.}{2013}]{LeeHI13}
{Lee} J.-M.,  {Heng} K.,   {Irwin} P.~G.~J.,  2013, \mn@doi [\apj]
  {10.1088/0004-637X/778/2/97}, \href
  {http://adsabs.harvard.edu/abs/2013ApJ...778...97L} {778, 97}

\bibitem[\protect\citeauthoryear{{Lewis}}{{Lewis}}{1969}]{Lew69}
{Lewis} J.~S.,  1969, \mn@doi [\icarus] {10.1016/0019-1035(69)90094-3}, \href
  {http://adsabs.harvard.edu/abs/1969Icar...10..393L} {10, 393}

\bibitem[\protect\citeauthoryear{{Lewis}, {Showman}, {Fortney}, {Marley},
  {Freedman}  \& {Lodders}}{{Lewis} et~al.}{2010}]{LewSF2010}
{Lewis} N.~K.,  {Showman} A.~P.,  {Fortney} J.~J.,  {Marley} M.~S.,  {Freedman}
  R.~S.,   {Lodders} K.,  2010, \mn@doi [\apj] {10.1088/0004-637X/720/1/344},
  \href {http://adsabs.harvard.edu/abs/2010ApJ...720..344L} {720, 344}

\bibitem[\protect\citeauthoryear{{Line} et~al.,}{{Line} et~al.}{2013}]{LinWZ13}
{Line} M.~R.,  et~al., 2013, \mn@doi [\apj] {10.1088/0004-637X/775/2/137},
  \href {http://adsabs.harvard.edu/abs/2013ApJ...775..137L} {775, 137}

\bibitem[\protect\citeauthoryear{{Line}, {Knutson}, {Wolf}  \& {Yung}}{{Line}
  et~al.}{2014}]{LinKW14}
{Line} M.~R.,  {Knutson} H.,  {Wolf} A.~S.,   {Yung} Y.~L.,  2014, \mn@doi
  [\apj] {10.1088/0004-637X/783/2/70}, \href
  {http://adsabs.harvard.edu/abs/2014ApJ...783...70L} {783, 70}

\bibitem[\protect\citeauthoryear{{Lodders}}{{Lodders}}{2004}]{Lod04}
{Lodders} K.,  2004, \mn@doi [\apj] {10.1086/421970}, \href
  {http://adsabs.harvard.edu/abs/2004ApJ...611..587L} {611, 587}

\bibitem[\protect\citeauthoryear{{Lodders}}{{Lodders}}{2010}]{Lod10}
{Lodders} K.,  2010, {Exoplanet Chemistry}.
p.~157, \mn@doi{10.1002/9783527629763.ch8}

\bibitem[\protect\citeauthoryear{{Lodders}, {Palme}  \& {Gail}}{{Lodders}
  et~al.}{2009}]{Lodders2009}
{Lodders} K.,  {Palme} H.,   {Gail} H.-P.,  2009, Landolt B{\"o}rnstein, \href
  {http://adsabs.harvard.edu/abs/2009LanB...4B...44L} {}

\bibitem[\protect\citeauthoryear{{Madhusudhan}}{{Madhusudhan}}{2012}]{Mad12}
{Madhusudhan} N.,  2012, \mn@doi [\apj] {10.1088/0004-637X/758/1/36}, \href
  {http://adsabs.harvard.edu/abs/2012ApJ...758...36M} {758, 36}

\bibitem[\protect\citeauthoryear{{Madhusudhan} \& {Seager}}{{Madhusudhan} \&
  {Seager}}{2009}]{MadS09}
{Madhusudhan} N.,  {Seager} S.,  2009, \mn@doi [\apj]
  {10.1088/0004-637X/707/1/24}, \href
  {http://adsabs.harvard.edu/abs/2009ApJ...707...24M} {707, 24}

\bibitem[\protect\citeauthoryear{{Madhusudhan} et~al.,}{{Madhusudhan}
  et~al.}{2011a}]{MadHS11}
{Madhusudhan} N.,  et~al., 2011a, \mn@doi [\nat] {10.1038/nature09602}, \href
  {http://adsabs.harvard.edu/abs/2011Natur.469...64M} {469, 64}

\bibitem[\protect\citeauthoryear{{Madhusudhan}, {Mousis}, {Johnson}  \&
  {Lunine}}{{Madhusudhan} et~al.}{2011b}]{MadMJ11}
{Madhusudhan} N.,  {Mousis} O.,  {Johnson} T.~V.,   {Lunine} J.~I.,  2011b,
  \mn@doi [\apj] {10.1088/0004-637X/743/2/191}, \href
  {http://adsabs.harvard.edu/abs/2011ApJ...743..191M} {743, 191}

\bibitem[\protect\citeauthoryear{{Madhusudhan}, {Bitsch}, {Johansen}  \&
  {Eriksson}}{{Madhusudhan} et~al.}{2017}]{MadBJ17}
{Madhusudhan} N.,  {Bitsch} B.,  {Johansen} A.,   {Eriksson} L.,  2017, \mn@doi
  [\mnras] {10.1093/mnras/stx1139}, \href
  {http://adsabs.harvard.edu/abs/2017MNRAS.469.4102M} {469, 4102}

\bibitem[\protect\citeauthoryear{{M{\'a}rquez-Neila}, {Fisher}, {Sznitman}  \&
  {Heng}}{{M{\'a}rquez-Neila} et~al.}{2018}]{MarFS18}
{M{\'a}rquez-Neila} P.,  {Fisher} C.,  {Sznitman} R.,   {Heng} K.,  2018,
  \mn@doi [Nature Astronomy] {10.1038/s41550-018-0504-2}, \href
  {http://adsabs.harvard.edu/abs/2018NatAs.tmp...81M} {}

\bibitem[\protect\citeauthoryear{{Mayne} et~al.,}{{Mayne}
  et~al.}{2014}]{MayBA14}
{Mayne} N.~J.,  et~al., 2014, \mn@doi [\aap] {10.1051/0004-6361/201322174},
  \href {http://adsabs.harvard.edu/abs/2014A%26A...561A...1M} {561, A1}

\bibitem[\protect\citeauthoryear{{Mendon{\c c}a}, {Grimm}, {Grosheintz}  \&
  {Heng}}{{Mendon{\c c}a} et~al.}{2016}]{MenGG16}
{Mendon{\c c}a} J.~M.,  {Grimm} S.~L.,  {Grosheintz} L.,   {Heng} K.,  2016,
  \mn@doi [\apj] {10.3847/0004-637X/829/2/115}, \href
  {http://adsabs.harvard.edu/abs/2016ApJ...829..115M} {829, 115}

\bibitem[\protect\citeauthoryear{{Mendon{\c c}a}, {Tsai}, {Matej}, {Grimm}  \&
  {Heng}}{{Mendon{\c c}a} et~al.}{2018}]{MenTM18}
{Mendon{\c c}a} J.~M.,  {Tsai} S.-M.,  {Matej} M.,  {Grimm} S.~L.,   {Heng} K.,
   2018, preprint, \href {http://adsabs.harvard.edu/abs/2018arXiv180800501M} {}
  (\mn@eprint {arXiv} {1808.00501})

\bibitem[\protect\citeauthoryear{{Menou}}{{Menou}}{2012}]{Men12}
{Menou} K.,  2012, \mn@doi [\apjl] {10.1088/2041-8205/744/1/L16}, \href
  {http://adsabs.harvard.edu/abs/2012ApJ...744L..16M} {744, L16}

\bibitem[\protect\citeauthoryear{{Menou} \& {Rauscher}}{{Menou} \&
  {Rauscher}}{2009}]{MenR09}
{Menou} K.,  {Rauscher} E.,  2009, \mn@doi [\apj]
  {10.1088/0004-637X/700/1/887}, \href
  {http://adsabs.harvard.edu/abs/2009ApJ...700..887M} {700, 887}

\bibitem[\protect\citeauthoryear{{Miguel}}{{Miguel}}{2019}]{Mig19}
{Miguel} Y.,  2019, \mn@doi [\mnras] {10.1093/mnras/sty2803}, \href
  {http://adsabs.harvard.edu/abs/2019MNRAS.482.2893M} {482, 2893}

\bibitem[\protect\citeauthoryear{{Miller-Ricci Kempton}, {Zahnle}  \&
  {Fortney}}{{Miller-Ricci Kempton} et~al.}{2012}]{MilZF12}
{Miller-Ricci Kempton} E.,  {Zahnle} K.,   {Fortney} J.~J.,  2012, \mn@doi
  [\apj] {10.1088/0004-637X/745/1/3}, \href
  {http://adsabs.harvard.edu/abs/2012ApJ...745....3M} {745, 3}

\bibitem[\protect\citeauthoryear{{Molli{\`e}re}, {van Boekel}, {Dullemond},
  {Henning}  \& {Mordasini}}{{Molli{\`e}re} et~al.}{2015}]{MolvD15}
{Molli{\`e}re} P.,  {van Boekel} R.,  {Dullemond} C.,  {Henning} T.,
  {Mordasini} C.,  2015, \mn@doi [\apj] {10.1088/0004-637X/813/1/47}, \href
  {http://adsabs.harvard.edu/abs/2015ApJ...813...47M} {813, 47}

\bibitem[\protect\citeauthoryear{{Mordasini}, {van Boekel}, {Molli{\`e}re},
  {Henning}  \& {Benneke}}{{Mordasini} et~al.}{2016}]{MorvM16}
{Mordasini} C.,  {van Boekel} R.,  {Molli{\`e}re} P.,  {Henning} T.,
  {Benneke} B.,  2016, \mn@doi [\apj] {10.3847/0004-637X/832/1/41}, \href
  {http://adsabs.harvard.edu/abs/2016ApJ...832...41M} {832, 41}

\bibitem[\protect\citeauthoryear{{Moses} et~al.,}{{Moses}
  et~al.}{2011}]{Moses2011}
{Moses} J.~I.,  et~al., 2011, \mn@doi [\apj] {10.1088/0004-637X/737/1/15},
  \href {http://adsabs.harvard.edu/abs/2011ApJ...737...15M} {737, 15}

\bibitem[\protect\citeauthoryear{{Moses}, {Madhusudhan}, {Visscher}  \&
  {Freedman}}{{Moses} et~al.}{2013a}]{Moses2013}
{Moses} J.~I.,  {Madhusudhan} N.,  {Visscher} C.,   {Freedman} R.~S.,  2013a,
  \mn@doi [\apj] {10.1088/0004-637X/763/1/25}, \href
  {http://adsabs.harvard.edu/abs/2013ApJ...763...25M} {763, 25}

\bibitem[\protect\citeauthoryear{{Moses} et~al.,}{{Moses}
  et~al.}{2013b}]{MosLV13}
{Moses} J.~I.,  et~al., 2013b, \mn@doi [\apj] {10.1088/0004-637X/777/1/34},
  \href {http://adsabs.harvard.edu/abs/2013ApJ...777...34M} {777, 34}

\bibitem[\protect\citeauthoryear{{Nissen}}{{Nissen}}{2013}]{Nis13}
{Nissen} P.~E.,  2013, \mn@doi [\aap] {10.1051/0004-6361/201321234}, \href
  {http://adsabs.harvard.edu/abs/2013A%26A...552A..73N} {552, A73}

\bibitem[\protect\citeauthoryear{{Nissen} \& {Gustafsson}}{{Nissen} \&
  {Gustafsson}}{2018}]{NisG18}
{Nissen} P.~E.,  {Gustafsson} B.,  2018, \mn@doi [\aapr]
  {10.1007/s00159-018-0111-3}, \href
  {http://adsabs.harvard.edu/abs/2018A%26ARv..26....6N} {26, 6}

\bibitem[\protect\citeauthoryear{{Nissen}, {Chen}, {Carigi}, {Schuster}  \&
  {Zhao}}{{Nissen} et~al.}{2014}]{NisCC14}
{Nissen} P.~E.,  {Chen} Y.~Q.,  {Carigi} L.,  {Schuster} W.~J.,   {Zhao} G.,
  2014, \mn@doi [\aap] {10.1051/0004-6361/201424184}, \href
  {http://adsabs.harvard.edu/abs/2014A%26A...568A..25N} {568, A25}

\bibitem[\protect\citeauthoryear{{{\"O}berg} \& {Bergin}}{{{\"O}berg} \&
  {Bergin}}{2016}]{ObeB16}
{{\"O}berg} K.~I.,  {Bergin} E.~A.,  2016, \mn@doi [\apjl]
  {10.3847/2041-8205/831/2/L19}, \href
  {http://adsabs.harvard.edu/abs/2016ApJ...831L..19O} {831, L19}

\bibitem[\protect\citeauthoryear{{{\"O}berg}, {Murray-Clay}  \&
  {Bergin}}{{{\"O}berg} et~al.}{2011}]{ObeMB11}
{{\"O}berg} K.~I.,  {Murray-Clay} R.,   {Bergin} E.~A.,  2011, \mn@doi [\apjl]
  {10.1088/2041-8205/743/1/L16}, \href
  {http://adsabs.harvard.edu/abs/2011ApJ...743L..16O} {743, L16}

\bibitem[\protect\citeauthoryear{{Oreshenko} et~al.,}{{Oreshenko}
  et~al.}{2017}]{OreLG17}
{Oreshenko} M.,  et~al., 2017, \mn@doi [\apjl] {10.3847/2041-8213/aa8acf},
  \href {http://adsabs.harvard.edu/abs/2017ApJ...847L...3O} {847, L3}

\bibitem[\protect\citeauthoryear{{Petigura} \& {Marcy}}{{Petigura} \&
  {Marcy}}{2011}]{PetM11}
{Petigura} E.~A.,  {Marcy} G.~W.,  2011, \mn@doi [\apj]
  {10.1088/0004-637X/735/1/41}, \href
  {http://adsabs.harvard.edu/abs/2011ApJ...735...41P} {735, 41}

\bibitem[\protect\citeauthoryear{{Piso}, {{\"O}berg}, {Birnstiel}  \&
  {Murray-Clay}}{{Piso} et~al.}{2015}]{PisOB15}
{Piso} A.-M.~A.,  {{\"O}berg} K.~I.,  {Birnstiel} T.,   {Murray-Clay} R.~A.,
  2015, \mn@doi [\apj] {10.1088/0004-637X/815/2/109}, \href
  {http://adsabs.harvard.edu/abs/2015ApJ...815..109P} {815, 109}

\bibitem[\protect\citeauthoryear{{Seager}, {Richardson}, {Hansen}, {Menou},
  {Cho}  \& {Deming}}{{Seager} et~al.}{2005}]{SeaRH05}
{Seager} S.,  {Richardson} L.~J.,  {Hansen} B.~M.~S.,  {Menou} K.,  {Cho}
  J.~Y.-K.,   {Deming} D.,  2005, \mn@doi [\apj] {10.1086/444411}, \href
  {http://adsabs.harvard.edu/abs/2005ApJ...632.1122S} {632, 1122}

\bibitem[\protect\citeauthoryear{{Showman}, {Fortney}, {Lian}, {Marley},
  {Freedman}, {Knutson}  \& {Charbonneau}}{{Showman}
  et~al.}{2009}]{Showman2009}
{Showman} A.~P.,  {Fortney} J.~J.,  {Lian} Y.,  {Marley} M.~S.,  {Freedman}
  R.~S.,  {Knutson} H.~A.,   {Charbonneau} D.,  2009, \mn@doi [\apj]
  {10.1088/0004-637X/699/1/564}, \href
  {http://adsabs.harvard.edu/abs/2009ApJ...699..564S} {699, 564}

\bibitem[\protect\citeauthoryear{{Stevenson}, {Bean}, {Madhusudhan}  \&
  {Harrington}}{{Stevenson} et~al.}{2014}]{SteBM14}
{Stevenson} K.~B.,  {Bean} J.~L.,  {Madhusudhan} N.,   {Harrington} J.,  2014,
  \mn@doi [\apj] {10.1088/0004-637X/791/1/36}, \href
  {http://adsabs.harvard.edu/abs/2014ApJ...791...36S} {791, 36}

\bibitem[\protect\citeauthoryear{{Teske}, {Cunha}, {Smith}, {Schuler}  \&
  {Griffith}}{{Teske} et~al.}{2014}]{TesCS14}
{Teske} J.~K.,  {Cunha} K.,  {Smith} V.~V.,  {Schuler} S.~C.,   {Griffith}
  C.~A.,  2014, \mn@doi [\apj] {10.1088/0004-637X/788/1/39}, \href
  {http://adsabs.harvard.edu/abs/2014ApJ...788...39T} {788, 39}

\bibitem[\protect\citeauthoryear{{Tremblin}, {Amundsen}, {Mourier}, {Baraffe},
  {Chabrier}, {Drummond}, {Homeier}  \& {Venot}}{{Tremblin}
  et~al.}{2015}]{Tremblin2015}
{Tremblin} P.,  {Amundsen} D.~S.,  {Mourier} P.,  {Baraffe} I.,  {Chabrier} G.,
   {Drummond} B.,  {Homeier} D.,   {Venot} O.,  2015, \mn@doi [\apjl]
  {10.1088/2041-8205/804/1/L17}, \href
  {http://adsabs.harvard.edu/abs/2015ApJ...804L..17T} {804, L17}

\bibitem[\protect\citeauthoryear{{Tremblin}, {Amundsen}, {Chabrier}, {Baraffe},
  {Drummond}, {Hinkley}, {Mourier}  \& {Venot}}{{Tremblin}
  et~al.}{2016}]{Tremblin2016}
{Tremblin} P.,  {Amundsen} D.~S.,  {Chabrier} G.,  {Baraffe} I.,  {Drummond}
  B.,  {Hinkley} S.,  {Mourier} P.,   {Venot} O.,  2016, \mn@doi [\apjl]
  {10.3847/2041-8205/817/2/L19}, \href
  {http://adsabs.harvard.edu/abs/2016ApJ...817L..19T} {817, L19}

\bibitem[\protect\citeauthoryear{{Tsai}, {Lyons}, {Grosheintz}, {Rimmer},
  {Kitzmann}  \& {Heng}}{{Tsai} et~al.}{2017}]{TsaLG2017}
{Tsai} S.-M.,  {Lyons} J.~R.,  {Grosheintz} L.,  {Rimmer} P.~B.,  {Kitzmann}
  D.,   {Heng} K.,  2017, \mn@doi [\apjs] {10.3847/1538-4365/228/2/20}, \href
  {http://adsabs.harvard.edu/abs/2017ApJS..228...20T} {228, 20}

\bibitem[\protect\citeauthoryear{{Tsai}, {Kitzmann}, {Lyons}, {Mendon{\c c}a},
  {Grimm}  \& {Heng}}{{Tsai} et~al.}{2018}]{TsaKL18}
{Tsai} S.-M.,  {Kitzmann} D.,  {Lyons} J.~R.,  {Mendon{\c c}a} J.,  {Grimm}
  S.~L.,   {Heng} K.,  2018, \mn@doi [\apj] {10.3847/1538-4357/aac834}, \href
  {http://adsabs.harvard.edu/abs/2018ApJ...862...31T} {862, 31}

\bibitem[\protect\citeauthoryear{{Venot}, {H{\'e}brard}, {Ag{\'u}ndez}, {Decin}
   \& {Bounaceur}}{{Venot} et~al.}{2015}]{Venot2015}
{Venot} O.,  {H{\'e}brard} E.,  {Ag{\'u}ndez} M.,  {Decin} L.,   {Bounaceur}
  R.,  2015, \mn@doi [\aap] {10.1051/0004-6361/201425311}, \href
  {http://adsabs.harvard.edu/abs/2015A%26A...577A..33V} {577, A33}

\bibitem[\protect\citeauthoryear{{Wakeford} et~al.,}{{Wakeford}
  et~al.}{2018}]{WakSD18}
{Wakeford} H.~R.,  et~al., 2018, \mn@doi [\aj] {10.3847/1538-3881/aa9e4e},
  \href {http://adsabs.harvard.edu/abs/2018AJ....155...29W} {155, 29}

\bibitem[\protect\citeauthoryear{{Waldmann}, {Tinetti}, {Rocchetto}, {Barton},
  {Yurchenko}  \& {Tennyson}}{{Waldmann} et~al.}{2015}]{WalTR15}
{Waldmann} I.~P.,  {Tinetti} G.,  {Rocchetto} M.,  {Barton} E.~J.,  {Yurchenko}
  S.~N.,   {Tennyson} J.,  2015, \mn@doi [\apj] {10.1088/0004-637X/802/2/107},
  \href {http://adsabs.harvard.edu/abs/2015ApJ...802..107W} {802, 107}

\bibitem[\protect\citeauthoryear{{Walsh}, {Nomura}  \& {van Dishoeck}}{{Walsh}
  et~al.}{2015}]{WalNv15}
{Walsh} C.,  {Nomura} H.,   {van Dishoeck} E.,  2015, \mn@doi [\aap]
  {10.1051/0004-6361/201526751}, \href
  {http://adsabs.harvard.edu/abs/2015A%26A...582A..88W} {582, A88}

\bibitem[\protect\citeauthoryear{{Woitke}, {Helling}, {Hunter}, {Millard},
  {Turner}, {Worters}, {Blecic}  \& {Stock}}{{Woitke} et~al.}{2018}]{WoiHH18}
{Woitke} P.,  {Helling} C.,  {Hunter} G.~H.,  {Millard} J.~D.,  {Turner} G.~E.,
   {Worters} M.,  {Blecic} J.,   {Stock} J.~W.,  2018, \mn@doi [\aap]
  {10.1051/0004-6361/201732193}, \href
  {http://adsabs.harvard.edu/abs/2018A%26A...614A...1W} {614, A1}


\end{thebibliography}





\bsp	
\label{lastpage}
\end{document}